\documentclass[10pt,a4paper]{article}
\usepackage[utf8]{inputenc}
\usepackage{amsmath}
\usepackage{amsfonts}
\usepackage{amssymb}
\usepackage{authblk}
\usepackage{graphicx}
\usepackage{chngcntr}
\usepackage{amsthm}

\newcommand{\keywordname}[1]{\bf Keywords: \normalfont}
\newcommand{\keywords}[1]{\par\addvspace\baselineskip
	\noindent\keywordname\enspace\ignorespaces#1}

\author[1]{James Gil de Lamadrid}
\affil[1]{Department of Computer Science\\Bowie State University\\14000 Jericho Pk. Rd\\Bowie, MD 20715\\ \texttt{jgildelamadrid@bowiestate.edu}}

\author[2] {Seonho Choi}
\affil[2]{Department of Computer Science\\Bowie State University\\14000 Jericho Pk. Rd\\Bowie, MD 20715\\
\texttt{schoi@bowiestate.edu}}
\title{Hardware Watermarking for Finite State Machines, with Symmetric Circuit Encryption}
\begin{document}
\maketitle
\counterwithout{figure}{section}
\theoremstyle{definition}
\newtheorem{definition}{Definition}
\theoremstyle{theorem}
\newtheorem{theorem}{Theorem}
\theoremstyle{property}
\newtheorem{property}{Property}
\theoremstyle{lemma}
\newtheorem{lemma}{Lemma}

\begin{abstract}
Putting a watermark into digital circuitry has its own set of challenges.  Creating a secure watermark in printed matter usually involves including graphics that are difficult to reproduce.  In circuitry, including additional circuitry that is hard to produce, one must contend with the prospect of increasing power consumption of the circuit, decreasing the speed of the circuit, and introducing watermark circuitry that is easily reproduced. 
\paragraph{}
In this paper we present a watermark method for sequential circuitry.  It allows for several degrees of calibration, allowing the user to tune the complexity of the watermark to requirements of speed, power, and efficacy.  Our method uses an encryption technique to introduce secrets about the watermark circuit, at several levels.  It also employs a boundary scan testing protocol as a means to protect the watermark circuitry.
\paragraph{}
Our discussion starts by describing the different tools needed in our watermark scheme.  We then discuss the difficulty of the problem associated with cracking our watermark circuit.  This analysis shows that, with full implementation, our method can be made quite secure.
\keywords{sequential circuits, watermarking, hardware security}
\end{abstract}

\section{Introduction}
A manufacturer involved in digital circuit production sometimes faces the problem of determining if a given physical device is an implementation of the manufacturer's design, or a functional copy.  A functional copy is a circuit implementation that performs the same task as the copied circuit.  It may or may not be implemented from the same design specification as that of the copied circuit.  Possibly the copier may have reverse-engineered an implementation of the device, and developed their own specification from that work.
\paragraph{}
A similar problem exists in the world of printed matter, where a publisher might be interested in establishing ownership of a document.  In such situations, the publisher might include a watermark on the document.  The watermark makes it very difficult to copy the document, and preserve the watermark as imprinted.  In this way, the publisher can distinguish between a verbatim copy of the material it owns, and a copy of only the content of its material.
\paragraph{}
We should stop here and point out several properties of watermarking:
\begin{enumerate}
\item
The purpose of a watermark is not to prevent copying.  Rather, faithful facsimiles of a circuit are quit acceptable.
\item
A watermark does not have to be undetectable.  In printed matter the watermark is often quite visible.
\item \label{p1}
Watermarks should be difficult to copy without access to proprietary information.
\item \label{p2}
A watermark should be unobtrusive.  In printed matter the watermark should not overpower the actual content.  In circuitry, the watermark should not detract excessively from the functionally, nor the performance of the digital circuit.
\end{enumerate}
\section{Circuit Watermark Classification}
Watermarking a circuit inevitably involves changing the design of the circuit to incorporate some characteristic that can be detected, given a physical implementation of the device.  The previous statement is a little misleading, since circuitry may have several design specifications, each specification at a different level of abstraction.  Because of this, as described in Charbon \cite{Charbon} we might actually classify watermarking by its level of abstraction.
\begin{itemize}
\item
Gate Level Watermarking.
\item
FSM Watermarking
\item
RTL Watermarking.
\end{itemize}
We have concentrated on finite state machine (FSM) watermarking, because it has certain advantages over register transfer level (RTL) watermarking, and gate level watermarking.
\paragraph{}
Property \ref{p1} of watermarks indicates that it should be difficult to copy a watermark.  It is conceivably possible to analyze a gate level watermark by examining the visible patterns of circuitry on a chip, making it easier to determine the circuitry changes made for watermarking.  Similarly, for RTL descriptions,  the components of the watermark might also  leave visible signatures on the silicon.  The FSM watermark has the advantage, that the FSM level, an intermediate level, uses a model that is perhaps less visible from the silicon.  To reverse engineer an FSM level description involves working from registers, and combinational circuitry back up through the gate level, to the FSM level, a much more complex process than just directly interpreting visual data derived from the silicon.
\paragraph{}
A second classification for hardware watermarking, discussed by Nie,\cite{Nie} is based on the nature of the modifications made to the circuit.  Some typical modification are listed below.
\begin{itemize}
\item
\textit{Behavioral Modification}.  In behavioral modification, new behavior is added to the circuit.  Keeping in line with Property \ref{p2}, the new behavior must not interfere with the normal operation of the circuit.  Often what is done is to add additional circuitry to the original design, to allow an engineer to perform a watermark test sequence.  Control changes would also be necessary to switch between a test mode and a normal operation mode.
\item
\textit{Structural Modification}. In structural modification, the behavior of the circuit is not changed. However, the structure of the circuit is modified  to include a feature which can be detected, usually, by examining the chip wiring.  This type of modification corresponds, conceptually, with the type of watermarking used in cartography, where a street map might be modified to include a small, non-existent street. When doing this type of watermarking, care must be exercised to remain in line with Properties \ref{p1}, and \ref{p2}.
\item
\textit{Side-effect Modification}.  We can talk about the normal design behavior of a circuit.  This behavior consists of producing and manipulating digital signals.  But, there are other observable  physical behaviors of a circuit that can be manipulated to produce a recognizable watermark event.  For example a circuit can be modified to produce a power consumption spike for a certain input, as describd by Gil de Lamadrid \& Choi.\cite{Gildelamadrid}  Other properties that could, or have been used for water marking are timing anomalies, temperature variation, and in some cases where output is tolerant of variation, modulation of output amplitude.
\end{itemize}
\paragraph{}
In this work, we concentrate on behavioral modification.  There are advantages, and disadvantages to this focus. A major disadvantage of this approach is that, unless complicated, any added behavior could, with effort, be analyzed, allowing a copier to reverse engineer the watermarked circuit, and determine the nature of the watermark event.  On the other hand, an advantage of the behavior watermark is that it is relatively easy to introduce into the FSM level design.
\paragraph{}
A significant requirement of the behavioral watermark is the need to create a watermark that is complex, but not arbitrary.  That is, the watermark must be the result of a sequence of decisions.  Knowing the results of these decisions, makes the watermark circuity explainable, and a lack of this knowledge makes the circuit design nonsensical.  This sequence of decisions functions as a set of hidden secrets.
\section{Watermarking and Encryption}
Our model of FSM watermarking in motivated by symmetric encryption (see Delfs et al \cite{Delfs} for further discussion). 
We now show how we can use encryption to develop watermark resistant to reverse engineering.  
\subsection{The Basics of Encryption}
In symmetric encryption, we have a piece of information that we wish to protect.  We will call this information the \textit{artifact}.  To protect the artifact, we change it by applying an encryption transformation.  we will refer to this transformation as a \textit{hash}.  
\paragraph{}
Hash functions use a secret piece of information called a \textit{key}.  To be more precise, a hash function is a function $h:A \times K \rightarrow B$, where $A$ is the set of artifacts, $K$ is the set of possible keys, and $B$ is the set of encryption.  Since the encryption process is meant to protect the artifact from discovery, the hash function must disguise the artifact sufficiently to prevent easy deciphering.  At the same time, the hash function must be relatively easy to compute, to keep the encryption process feasible.
\paragraph{}
A fundamental of symmetric encryption is that an encryption of an artifact should be decipherable.  That is, there is a reverse hash function that, given an encrypted artifact, and the key value, outputs the original artifact.  In other words, a decryption function is a function $h^{-1}: B \times K \rightarrow A$, such that $h^{-1}(h(a, k), k) = a$, for $a \in A$.
\subsection{The Basics of Behavioral Watermarking}
Of course encryption is usually performed on artifacts that are integer values.  The question that now comes up is how can encryption be used in watermarking?  To answer that question we need to explain the basics of behavioral watermarking.
\paragraph{}
In our version of behavioral watermarking, circuitry is added to a circuit that enables a test engineer to perform a \textit{watermark test}, as is done in the work of Liang et al.\cite{Liang}  In a watermark test, the test engineer provides a sequence of input signals to the circuit.  With the watermark circuit present, this will cause the device to produce a sequence of test outputs.  The test engineer can determine if the watermark circuitry is present by comparing the output of the device to the expected watermark outputs.
\subsection{Viewing a Watermark as an Encryption}\label{p3}
We can view a watermark circuit as an encryption artifact.  Artifacts are normally numbers.  A watermark circuit, at the FSM level, represented as an adjacency matrix, is, in fact numeric, and quit amenable to encryption.  
\paragraph{}
Our goal is to make a watermark difficult to make sense of.  This hampers an effort to copy it without careful attention to detail.  We make the watermark appear to be nonsensical by applying transformations that are held in secret.  Each transformation obfuscates the actual watermark specification even more.
\paragraph{}
We might augment our repertory of techniques to include more than just transformation.  The more secrets, whether transformation or not, the more knowledge a copier must posses to make sense of the watermark.  We start our watermark construction, by creating a circuit that is derived from the host FSM.  That the watermark starts with a circuit derived from the original device is in itself a secret.  We call the derived circuit a \textit{reduction} (REDUX), because it is of reduced size, compared to the original host.
\paragraph{}
The process of creating a watermark, as described so far, consists of three steps.
\begin{enumerate}
\item
From the host FSM, create a REDUX.
\item
Encrypt the reduction using s key value.
\item
Package the host machine, and the encrypted REDUX as the final circuit.
\end{enumerate}

To test the watermark, the test engineer takes the encrypted REDUX, and decrypts it, yielding a functioning REDUX circuit for the host machine.  To decrypt the watermark circuit, a decryption machine is connected to the watermark machine.  This decryption machine transforms the outputs of the watermark FSM into the outputs of the full REDUX machine.
\paragraph{}
Notice that we have separated,or decomposed, the REDUX into two separate machines: the encryption, which is distributed with the host machine, and a decryption machine, that is available to the test engineer only.
\paragraph{}
In a test sequence, the test engineer would have at their disposal, a copy of the full REDUX, a copy of the decryption machine, and a copy of the test machine.  The test process would entail the following three steps.
\begin{enumerate}
\item
Connect the decryption machine to the test machine
\item
Run the test-decryption machine on a sequence of test inputs.  Simultaneously run the REDUX on the same test sequence
\item
Compare the output of the test-decryption machine with the output of the REDUX.  If the output is identical, then the evidence suggests that the test machine contains the watermark circuitry.  If not, the evidence suggest that the watermark is absent from the test machine.
\end{enumerate}
\section{Reducing the Size of an FSM}\label{pp3}
The FSM model for a sequential circuit can be quite large.  To form a watermark circuit out of the host FSM, it beneficial transform it into a graph that has fewer states.  This decreases the work required to develop the watermark, but more importantly, it decreases the size of the watermark circuit, making it less obtrusive.  The reduced graph, as mentioned in Section \ref{p3}, is called the REDUX.  A REDUX can be developed from the host machine in several ways.  We propose the use of a REDUX called a \textit{longest path reduction} (LPR).  We do this simply to obtain a simple reduction that can be used to demonstrate our method.
\subsection{Calculate the Connectivity Graph of the Host FSM}
\begin{figure} [h]
\centering
\includegraphics[scale=0.25]{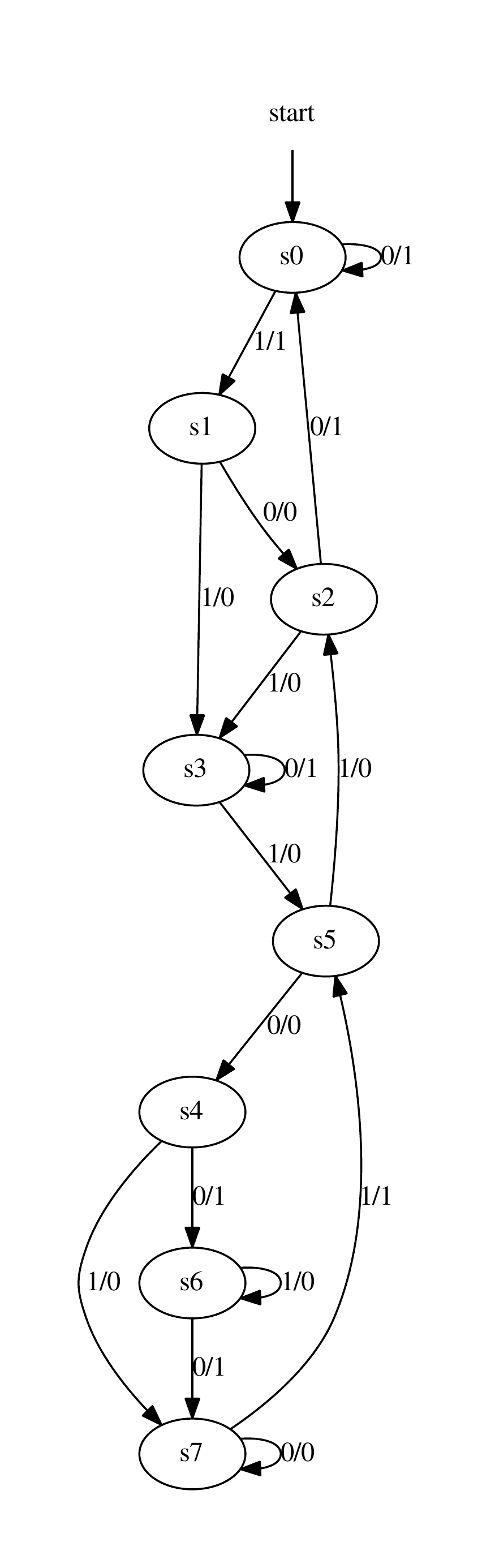}
\caption{Example Mealy machine.\label{f1}}
\end{figure}

\begin{figure} [h]
\centering
\includegraphics[scale=0.25]{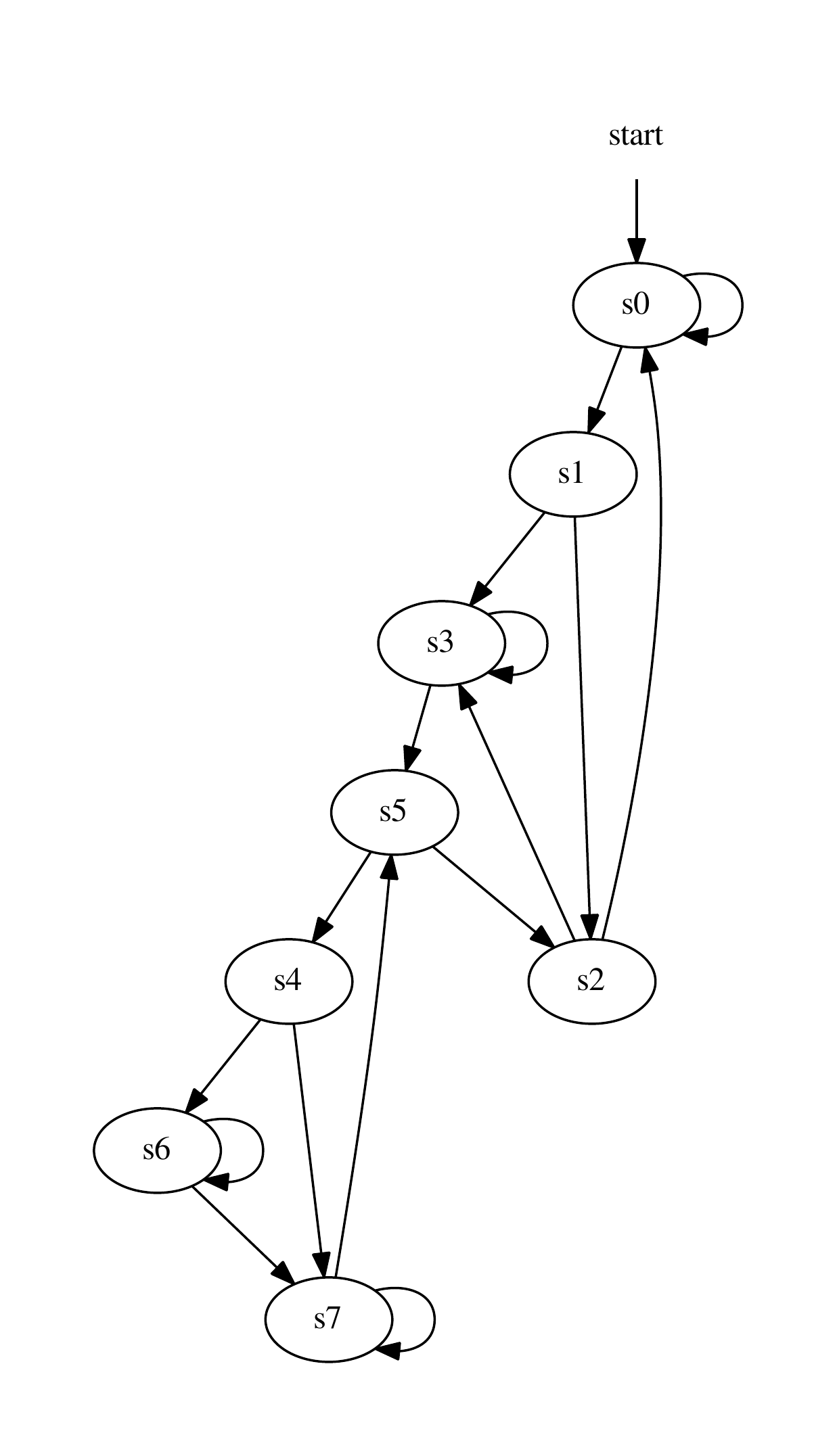}
\caption{Connectivity graph for the example.\label{f2}}
\end{figure}

When developing a watermark circuit for an FSM, we are looking for a circuit with a structure that depends on the original host machine, but it is not necessary that the watermark machine contains all information contained in the host.  Probably the most interesting information contained in an FSM description is its topology.  Intuitively, then, we might decide to dispose of information on the inputs and outputs of the host FSM.  This is the idea behind the \textit{connectivity graph} (CG) of an FSM.  The connectivity graph of an FSM is a \textit{rooted} digraph, where a rooted digraph is a digraph with a designated start vertex.  We now define the CG more formally.
\paragraph{}
Let $M = (S, T, I, P, O, s)$ be an FSM, where \\
\indent
$S$ is the set of states,\\
\indent
$I$ is a set of inputs,\\
\indent
$P$ is a set of outputs,\\
\indent
$s$ is a reset state,\\
\indent
$T$ is the transition function; $T : S \times I \rightarrow S$, and\\
\indent
$O$ is an output function; $O :  S \times I \rightarrow P$.\\
This definition corresponds to the definition of a general Mealy machine.  
\begin{definition} \textit{Connectivity Graph.}
The CG of $M$ is the rooted digraph $G_M = (V_M, E_M, s)$, where 
\begin{itemize}
\item
$V_M = S$ is the set of vertexes,
\item
$E_M$ is the set of edges of $G_M$, and
\item
$s$ is the root.
\end{itemize}
The set of edges, $E_M$, consists of the the edges induced by the transition function of the host, $T$, with disregard to any input or output.  More formally, $E_M$ is defined as a function $E_M : V_M \rightarrow V_M$ such that
\begin{equation}
E_M = \{(s_1 , s_2 ) | s_1, s_2 \in S \wedge \exists x \in I \wedge (s_1, x, s_2) \in T\}.
\end{equation}
\end{definition}
\paragraph{}
In Figure \ref{f1}, we show an example FSM, and in Figure \ref{f2} we show the corresponding CG.  Th graph shown has eight states.  More succinctly, we would say that it is an element of $FSM_8$.  In general, $FSM_n$ is the set of connectivity graphs with $n$ states.  
\paragraph{}
Notice that the connectivity graph has the same number of states as the host machine.  Using the CG in creating the watermark does not reduce the number of states, but rather the number of transitions.  Reducing the number of states is done by the next transformation.
\subsection{Calculate the LPR of the Host Machine}
Reducing the number of states in a CG, can be done by any number of transformations.  If trivial reductions are used,  one will end up with no reduction in states, or a reduction that does not characterize the CG. Therefor, a reduction is needed that  is much less trivial than these two extremes.
\paragraph{} 
The reduction that we use, as stated in Section \ref{pp3}, is called the \textit{longest path reduction}.  In brief, we characterize the CG with a subgraph which is the longest acyclic path from the start state.
\paragraph{}
Of course, finding the longest path in a general graph is an NP-Hard problem (see Garey \& Johnson\cite{Garey} for the reduction proof), and until it is shown that P = NP, we are more inclined to settle for an approximation.  So, our approximation to an actual longest path is the longest path that can be discovered with a depth-first search.  For a connected graph, this can be found in $O(v^2)$ time, where $v$ is the number of vertices in the graph (see McConnel \cite{Mcconnel}).
\paragraph{}
Another consideration concerning the LPR is that it is reasonable to desire control over the size of the LPR.  This is not possible with our current definition.  For instance, if the LPR has a length of $n$, but we would like the LPR to be length $k$, if $k \le n$, we have no problem, we can truncate the LPR to length $k$.  On the other hand, if $k > n$, we need a method of extending the LPR to length $k$.  A simple way to do this is by repeating the LPR to increase its length.
\paragraph{}
We now formally define an LPR of a CG.  We start by defining the longest path that can be discovered by depth first traversal (DFT) of the CG.
\begin{definition} \textit{Longest Simple Path.}
Let $G = \{V, E, v\}$ be a rooted digraph.  Let $\prec : V \times V \rightarrow \{0, 1\}$ be a total ordering of $V$.  Let $P$ be the set of all \textit{depth first traversal} (DFT) paths from the vertex $v$.  The ordering $\prec$ induces a total lexical ordering on the set $P$, $\prec^*$.  Therefor, there is a maximal element, $p_{max} \in P$.  $p_{max}$ can be referred to as the \textit{longest simple path} in $G$.
\end{definition}
Notice that although $\prec$ is arbitrary, if the states in the graph $G$ are represented by numbers, it would be intuitive to order the states by their state numbers.  Also notice that, since $\prec^*$ is lexical, it is an ordering of the paths in $P$ by their path lengths.
\paragraph{}
To adjust the length of the longest simple path to the desired value, we repeat and/or truncated it.  A slight problem is that repetition will cause vertexes to occur multiple times in the resulting vertex string.  This problem can be solved by renumbering the vertexes in the string.  
\begin{definition} \textit{Finite Repetition.}
 Let $p \in V^i$. Then $p^j \in V^{i \times j}$ is the vertex string formed by repeating the string $p$, $j$ times.
\end{definition}
\begin{definition} \textit{Renumbering.}
Let $p \in V^i$, where \\
\indent
$p = \langle \langle v_{r, c} : 1 \le c \le i \rangle : 1 \le r \le j \rangle$.\\
Since the string $p$ is a repeated string, we see that $\forall k_1, k_2, v_{k_1, c} = v_{k_2, c}$, and so it is appropriate to refer to this repeated vertex as just $V_c$. Let  $v_* = \max\limits_{1 \le c \le i}(v_c)$.
Let \\
\indent
$\gamma_{v_*}(v_c, r) = v_* \cdot (r - 1) + v_c$.\\
Then the function $\gamma_{v_*}$ defines a \textit{renumbering} of the vertexes in $p^j$.  That is,\\
\indent
$\gamma(p^j) = \langle \langle \gamma_{v_*}(v_c, r) : 1 \le c \le i \rangle : 1 \le r \le j \rangle$.
\end{definition}
\paragraph{}
Notice that $\gamma_{v_*}$ merely encodes the row number $r$ into the vertex number $v_c$, using a fairly standard radix encoding.  The function $\gamma_{v_*}$ has an inverting function defined as \\
\indent
$\gamma_{v_*}^{-1}(w) = v_{(w \mod v_*) + 1}$,\\
and the inverse of $\gamma$ would be\\
\indent
$\gamma^{-1}(\langle v_i : 1 \le k \le i \times j \rangle) = \langle \gamma_{v*}^{-1}(v_k) : 1 \le k \le i \times j \rangle$.
\begin{definition} \textit{Truncation.}
 Let $p \in V^i$.  Then $\tau_j(p) \in V^j$, for $0 \leq j \leq i$, is the string $p$ truncated to its first $j$ vertexes.
\end{definition}
\paragraph{}
We are now able to construct a string of vertexes of any desired length, $m$, out of a longest simple path of length $n$, using the function $\sigma_{n,m}$ defined as follows.
\begin{definition} \textit{Sized Longest Simple Path.}
Let $p = \langle v_k : 1 \le k \le n \rangle$.  Then \\
\indent
$\sigma_{m}(p) =  \tau_m(\gamma_{v_*}(p^j)) \in V^m$\\
where $j = \lceil {m \over n} \rceil$.
\end{definition}
Finally we can define the LPR, by taking the sized longest simple path, and converting it into an FSM.  In this definition, we use the notation $G_n$ to indicate the class of all rooted digraphs of size $n$.  That is, $G_n = \{G = (V, E, v) : \| V \| = n\}$.
\begin{definition} \textit{Longest Path Reduction.}
Let $p$ be the longest simple path in a graph $G$.
Let $\Sigma : G_n \rightarrow G_m$, such that \\
\indent
$\Sigma(G) = H$\\
where $G = (V, E, v) \in G_n$ is a CG, and $H = (V_{\sigma_m(p)}, E_{\sigma_m(p)}, v) \in G_m$, is also a CG.  In the above definition, $V_{\sigma_m(p)} = \{w  | w \in \sigma_m(p)\}$, and $E_{\sigma_m(p)} = \{(v_i, v_{i+1}) | \sigma_m(p) = (v_1, v_2, v_3, \ldots , v_k) \wedge 1 \leq i < k\}$.  Then the longest path reduction of $G$ is the CG $\Sigma(G)$.
\end{definition}
Intuitively, we form the LPR of a graph by truncating, and repeating its longest simple path, and then forming a CG that includes all of the vertexes and edges of the sized path.
\subsection{Use of the LPR}
\begin{figure} [h]
\centering
\includegraphics[scale=0.25]{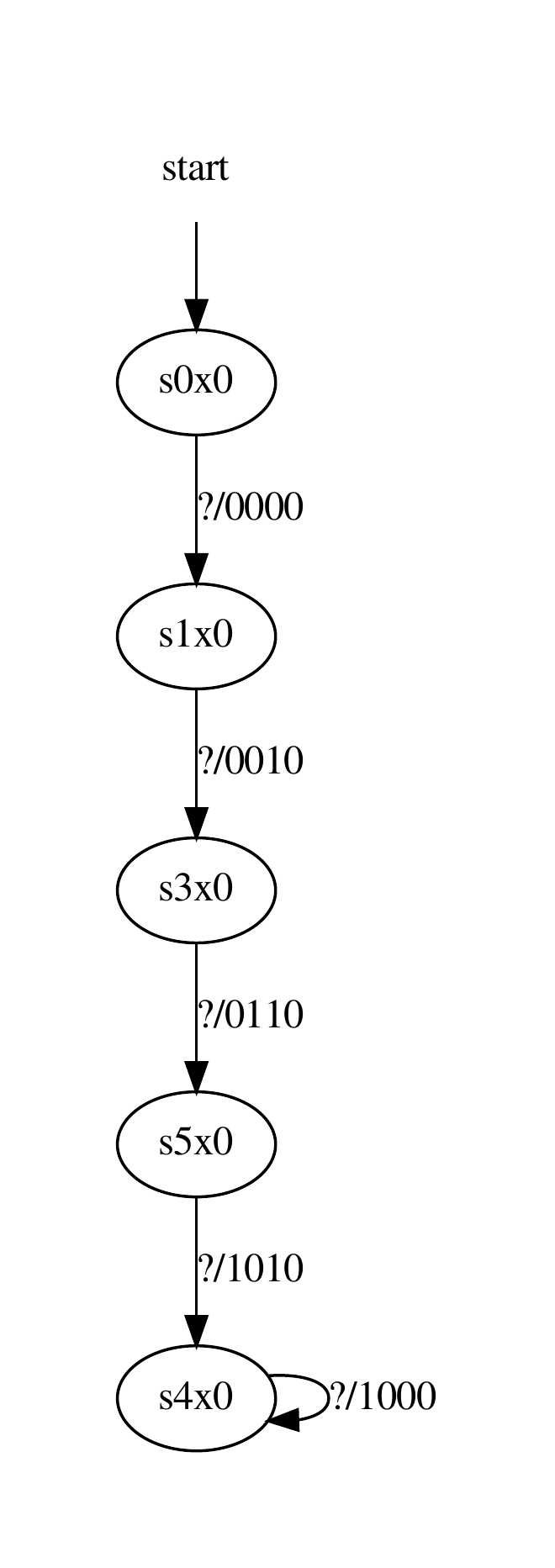}
\caption{LPR for the running example.\label{f3}}
\end{figure}

The LPR of an FSM, $M$, can be considered a ``distillation'' of the FSM $M$.  We use this distillation of the FSM as a \textit{signature} that characterizes the original host machine, to some degree.  Using the term \textit{signature} to describe the LPR of $M$, is with intent.  We now start to relate our watermarking procedure to encryption.  
\paragraph{}
The signature LPR is the circuit from which we will derive our watermark.  We derive the watermark circuit by encrypting the signature.
\paragraph{}
For the running example which we are presenting, Figure \ref{f3} shows the LPR of the original machine from Figure \ref{f1}.
\section{Symmetric Encryption of the REDUX}\label{s5}
In the next step of our procedure, we protect, or hide the signature machine.  That is, if we are using the LPR as the REDUX, we encrypt $\Sigma(G_M)$, where $G_M$ is the CG of the host machine $M$.
\paragraph{}
To symmetrically encrypt a signature machine $\Sigma(G_M)$, we can think of the encryption process as a transformation that uses a key value $\kappa$.  That is, we apply a function $\lambda$ to the signature, where $\lambda : Z^* \times G_m \rightarrow G_p$, and $\Sigma(G_M) \in G_m$.
\begin{definition} \textit{Machine Encryption.}\\
\indent
$\lambda(\kappa, \Sigma(G_M)) = \lambda_{\kappa,M}$\\
is called a an encryption of machine $M$, with key $\kappa$.
\end{definition}
\paragraph{}
We now have a definition of an encryption.  For a symmetric encryption, the function $\lambda$ must satisfy the \textit{symmetry property}.
\begin{definition}
\textit{Symetry Property.}
A function $\lambda$ satisfies the \textit{symmetry property} if and only if there exists a \textit{decryption function}, $\lambda^{-1} : Z^* \times G_p \rightarrow G_m$ such that
\begin{equation} \label{e2}
\lambda^{-1}(\kappa, \lambda(\kappa, G_M)) = G_M.
\end{equation}
\end{definition}

\paragraph{}
The nature of the function $\lambda$ can vary extensively.  We will concentrate on a couple of simple transformations:
\begin{itemize}
\item
Function Composition,
\item
Machine Composition.
\end{itemize}
\subsection{Function Composition}
In function Composition transformation, the encryption transformation, and the decryption transformation are viewed as function application.  The result is that the transformations are a litteral interpretation of the procedure described by Equation \ref{e2}.
\paragraph{}
We demonstrate the concept of functional composition with a fairly straight-forward implemention: matrix multiplication.  This is a technique based on the Hill Cipher.\cite{Stalling}  That is, we can think of $\lambda$ as a function that multiplies a matrix by a transformation matrix, and $\lambda^{-1}$ as a multiplication by the inverse of the transformation matrix.
\subsubsection{Function Composition with Matrix Multiplication}\label{s511}
To describe function composition transformation as matrix multiplication, we will first introduce some terms.  Let $G_M$ be a CG.  We denote its adjacency matrix representation as $\rho(G_M)$, where $\|\rho(G_M) \|= m \times m$, and $\rho^{-1}(A)$ denotes the CG corresponding to an adjacency matrix $A$.
\begin{definition} \textit{Encryption and Decryption Functions.}
Let $K$ be an orthogonal matrix, with $\| K \| = m \times m$.  Since $K$ is orthogonal, it has an inverse, $K^{T}$ such that \\
\indent
$\rho(G_M) \times K \times K^{T} = \rho(G_M)$.\\
We define the transformation function $\lambda$ as follows.
\begin{equation}\label{e3}
\lambda(K, G_M) = \rho^{-1}(\rho(G_M) \times K)
\end{equation}
\begin{equation}\label{e4}
\lambda^{-1}(K, G_M) = \rho^{-1}(\rho(G_M) \times K^{T}).
\end{equation}
\end{definition}
\paragraph{}
As can be seen by comparing Equations \ref{e3}, and \ref{e4} with the work presented at the beginning of Section \ref{s5}, we see that the matrix $K$ is being used as the key in the encryption.  A key, $\kappa$, is usually considered to be a number. However, we can extend this role to matrices, without loss of generality, since it is fairly easy to construct a lossless encoding of  a matrices into a numbers.
\subsubsection{Implementing Matrix Multiplication Encryption with the LPR} \label{s19}
\begin{figure} [h]
\centering
\includegraphics[scale=0.25]{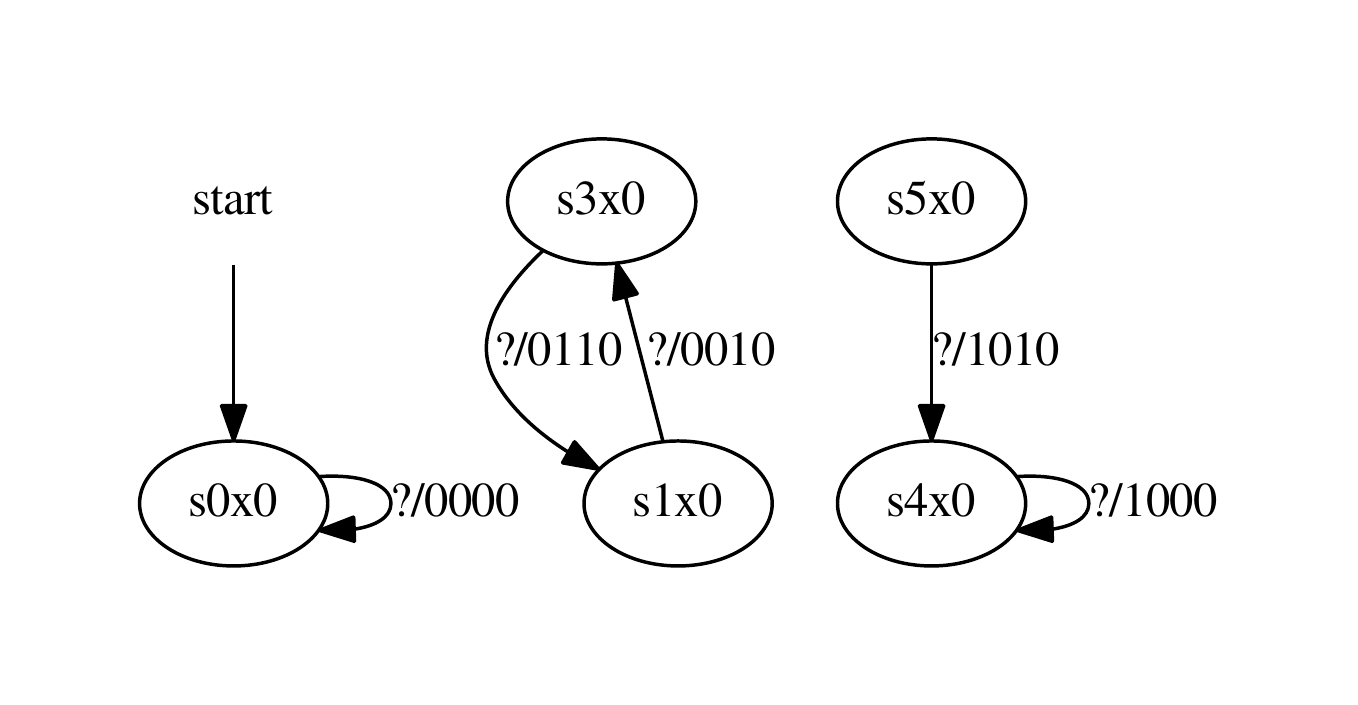}
\caption{Example matrix encrypted machine.\label{f4}}
\end{figure}

The question we now consider is how do you use the results on encryption with matrix multiplication to construct a watermark circuit from the LPR?  We will point out here that we have already shown how to encrypt a CG, in Section \ref{s511}.  Remember that the LPR of the host machine $M$ is, in fact, a CG, which we can refer to as $G_{LPR_M}$.  We can therefor encrypt $G_{LPR_M}$ into a CG which can be used to construct the FSM that will be used as the watermark circuit. 
\paragraph{}
We can use a standard method for creating a FSM from a CG.  This method is, as presented here, applied to connectivity graphs that have the same form as an LPR, but it can be generalized.  We call a CG with the same form as an LPR a \textit{linear graph}.  
\begin{definition} \textit{Linear Graph}  A CG $G = (V, E, s)$ is a linear graph if and only if, for all vertices $v \in V$, the out-degree of $v$ is 1.
\end{definition}
\paragraph{}
Recall that the key matrix $K$ used in matrix multiplication is orthogonal, and with Boolean values this implies that $K$ is the identity matrix, with the rows permuted in some fashion.  We call this type of matrix an \textit{identity permutation}.
\paragraph{}
What we now describe is a method of converting any linear graph into an FSM.  Since the encryption of the LPR, $\lambda(K, G_{LPR_M})$, is a linear graph, we can use this method to convert it into an FSM.  Further, we can use this conversion method as a standard technique to convert any CG into an FSM.
\paragraph{}
The FSM we build from a linear graph is a simple machine that simply outputs its current state, and receives no input.
\begin{definition}\textit{Standard CG Machine}.  The standard CG machine, for a connectivity graph $G = (V, E, s)$ is an FSM defined as 
\begin{equation}
\phi(G) = (V, E, \varnothing, V, O_G, s),
\end{equation}
where $O_G : E \rightarrow V$ is defined as\\
\indent
$O_G(u, w) = u$.
\end{definition}
To summarize the machine $\phi(G)$, it is a machine with the same vertexes as the graph $G$, its transitions consist of the edges of $G$, it has no input, its output are vertices, it outputs the source vertex on an edge, and its start vertex is the same as it is for $G$.
\paragraph{}
Figure \ref{f4} shows the encrypted machine for our running example.  As can be seen, It is topologically very distinct from the original LPR, being a non-linear disconnected FSM.
\subsubsection{Implementing Matrix Multiplication Decryption for the LPR}
\begin{figure} [h]
\centering
\includegraphics[scale=0.25]{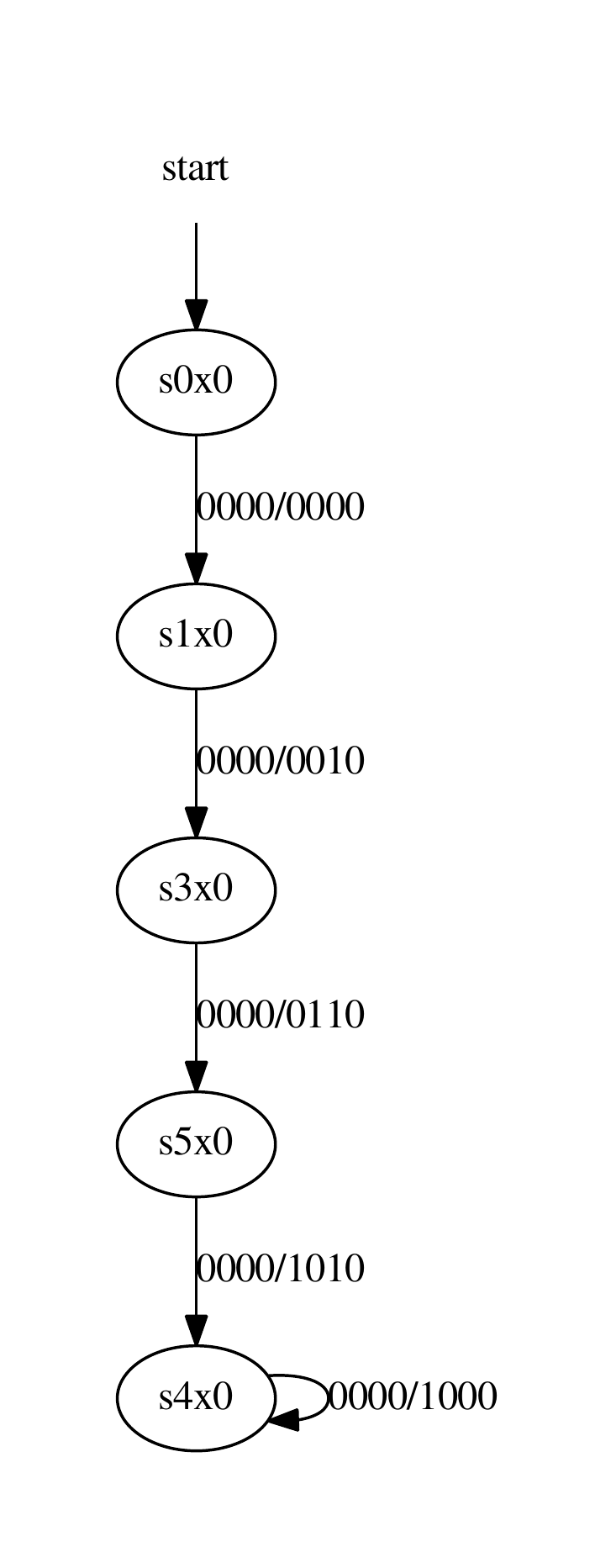}
\caption{Example matrix decryption machine.\label{f5}}
\end{figure}

A problem with decryption is that we do not have access to the encrypted watermark machine, defined in Equation \ref{e3}.  This machine is hidden inside of the package that contains both the host machine and $\lambda (K, G_M)$, and so it is impossible to actually multiply the watermark machine by $K^T$, as described in Equation \ref{e4}.  It is fortunate, though, that the machine $\lambda(K, G_M)$ outputs its state.  So, instead of multiplying the encrypted machine by the decryption matrix, we can build a decryption machine that simulates the original machine, $G_M$, given correct output from  $\lambda(K,G_M)$.  This new machine can be composed with the watermark machine.  By simulate, we mean that given the same inputs, the decryption machine will produce the same outputs as the non-encrypted machine.
\paragraph{}
To build the decryption machine we use the technique of tracing, used when converting an NFA into a DFA.\cite{Lintz}  That is, we trace the machine $LPR_M$, and $\lambda(K, LPR_M)$, simultaneously.  Each machine is given the appropriate input, at each time step.  As we trace the wo machines we keep track of the current states of the two machines.  This list of current states is then used to construct the decryption machine.
\paragraph{}
If the machine we are tracing is an LPR, the trace is significantly simplified. In particular, an LPR is a linear graph.  To be more descriptive, both the LPR, and the encrypted machine are linear graphs.  In particular, the LPR is a chain of states of length, let us say, $m$.
\paragraph{}
We can define our trace more formally.  Let $\langle (u_k, v_k) \rangle$ be a  time sequence, based on time step $k$, where $u_k \in LPR_M.V \wedge v_k \in (\lambda(K, LPR_M)).V$. (The dot notation, $G.V$, indicates a component of the CG $G = (V, E, s)$.) We can define our series, inductively, as follows.
\begin{equation}
u_k = 
\begin{cases}
LPR_M.s, & k = 0 \\
 u, & k > 0 \wedge (u_{k-1}, u) \in LPR_M.E
\end{cases}
\end{equation}
\begin{equation}
v_k =
\begin{cases}
(\lambda(K, LPR_M)).s, & k = 0\\
v, & k > 0 \wedge (v_{k-1}, v) \in (\lambda(K, LPR_M)).E
\end{cases}
\end{equation}
That is, at each time step, $k$, for both machines, we follow the edge from the state for the previous time step, $k-1$ to the new state.
\paragraph{}
Another tool we will need is concerned with generating the output of the decryption machine.  The decryption machine receives the output of the encrypted machine, $\lambda(K, LPR_M)$, as its input.  The decryption machine, $\lambda^{-1}(K, LPR_M)$, then must output the state of the original machine, $LPR_M$.  This is done by taking the input from the encrypted machine and ``multiplying'' it by the matrix $K^T$.  We now define this multiplication process in more detail.
\begin{definition}
\textit{Decryption Machine.}
The decryption machine is defined as
\begin{equation}
\lambda^{-1}(K, LPR_M) = (V, T', V,V, O', s),
\end{equation}
where $V = LPR_M.V$, $s = LPR_M.s$, $T' : V \times V \rightarrow V$ is defined as\\
\indent
$T' = \{(v_k, u_k, v_{k+1}) | 0 \le k < \|LPR_M.V\| - 1$,\\
and $O' : V \times V \times V \rightarrow V$ is defined as\\
\indent
$O'(v_k, u_k, v_k) = v_k$
\end{definition}
\paragraph{}
Essentially, the decryption machine mimics the original machine $LPR_M$, but only moves from one state to another if it receives the correct input from the encrypted machine $\lambda(K, LPR_M)$.  The decryption machine outputs its current state, as usual for a CG machine.  We show the decryption machine for our running example in Figure \ref{f5}.
\subsection{Machine Composition}\label{s22}
In machine composition, the emphasis is not so much on transforming the calculation performed by a machine, but more decomposing the machine into two pieces, each of the pieces finite state machines.  In this view, unlike in function composition, transformation is not semantic in nature, but can be purely arbitrary, and syntactic, in a sense.  This type of transformation is based on FSM partitioning theory (see Lee \& Yannakakis\cite{Lee}).
\paragraph{}
The strategy involved in machine composition is the same as that for function composition.  Just as in function composition, we break an original FSM into two machines.  Just as in function composition, one machine is used as the encrypted machine, and one is used as the decryption machine.  We present two techniques as examples of machine composition.  Both techniques are related, as we will discuss.
\begin{itemize}
\item
\textit{Optimal Machine Decomposition},  In this technique we enumerate all possible decompositions of an FSM into two submachines, and choose a smallest possible decomposition.  Optimality is achieved by minimizing the number of states in the two composite machines.
\item
\textit{Approximate Machine Decomposition}.  In this technique we heuristically construct a machine that has a particular fixed structure.  The quality of the constructed machine will be determined by the size of the composite machine.
\end{itemize}
\paragraph{}
In terms of advantages and disadvantages, an optimal decomposition produces a pair of machines with a small number of states, making the watermark circuit smaller, with the usual advantages of small circuitry.  On the other hand finding the optimal circuit is NP-Complete, and requires a significant effort.  This leads one to look for a good approximation, which we have investigated.  The approximation we have fond is actually an optimal solution.  Using this approximation then eliminates the need to search for an optimal solution, and also provides a small watermark circuit.  The approximation, however, may have the disadvantages discussed in Section \ref{sIA}.
\subsection{Optimal Machine Decomposition}
As mentioned, much of this work is rooted in the theory that has been worked out concerning FSM decomposition.\cite{Lee}\cite{Desai}  We present only enough of this work to explain our own.
\paragraph{}
The goal, in machine decomposition is to split a given FSM into two separate machines.  These two machines, when composed together are capable of simulating the original machine.  Now, this task in and of itself is relatively simple.  We can split a machine $M$ into two machines, $\mathbf{1}$, and $\mathbf{0}$, where $\mathbf{0} = M$, and $\mathbf{1}$ is an FSM with only one state.  This decomposition is termed \textit{trivial}, for trivially obvious reasons.  So, we modify our problem; we wish to decompose the machine $m$ into two nontrivial machines.  Also we would also stipulate that to correspond to our intuitive notion of what a decomposition is, the two machines should be smaller than $M$.  That is, if $M$ is decomposed into machines $M_D$, and $M_I$, then $\|M_D.V\| \le \|M.V\| \wedge \|M_I.V\| \le \|M.V\|$.
\paragraph{}
Work has been done on mostly two types of decomposition.\cite{Lee}  The difference between the two types is the nature of how the decomposed machines are composed together to produce the simulator of the original machine.
\begin{itemize}
\item
\textit{Serial/Cascade Decomposition}. In this type of decomposition a machine $M$ is decomposed into two machines $M_I$, and $M_D$.  The machine $M_I$ handles the input of $M$, and its output is processed by $M_D$, which, in turn, produces the output of $M$, implementing a pipeline structure.  More specifically, if $M = (V, T, I, P, O, s)$, then $M_I = (V_I, T_I, I, I \times V_I, O_I, s_I)$, and $M_D = (V_D, T_D, I \times V_I, P, O_D, s_D)$.  The machine $M_D$ is called the \textit{dependent machine}, and $M_I$ is called the \textit{independent machine}.
\item
\textit{Parallel Decomposition}. In this type of decomposition, there are actually not two decomposed machines, but actually three.  We will call these machines the two \textit{factor machines}, and the \textit{glue machine}.  In this scheme, the two factor machines run independent of each other, each receiving the input for $M$.  Each factor machine outputs its state.  The glue machine receives the states from the two factor machines, and computes the output necessary to simulate $M$.  More precisely, if  $M = (V, T, I, P, O, s)$, then $M$ is decomposed into three machines, $M_1$, $M_2$, and $M_G$ such that $M_1 = (V_1, T_1, I, V_1, O_1, s_1)$, $M_2 = (V_2, T_2, I, V_2, O_2, s_2)$ and $M_G = (V_G, T_G, I \times V_1 \times V_2, P, O_G, s_G)$.
\end{itemize}
In our work we have concentrated on cascade decomposition, since it seems like a good fit for our scenario, where an encrypted machine is producing output that is interpreted by a decryption machine, in a cascade fashion.
\subsubsection{State Partitioning}
A substantial amount of work has been done on decomposing a machine by partitioning the states of the machine, described bu Lee \& Yannakakis.\cite{Lee}  This involves two operations, although the procedure used does not follow these operations as an algorithmic sequence.
\begin{enumerate}
\item
Create two machines, each one containing all of the states in the original machine.
\item
In each machine, group states together forming \textit{blocks}.  The states of the two machines, the dependent and independent machines, will be blocks of states form the original machine.  This procedure is very similar to the usual technique used in machine minimization,(see Imeida et al.\cite{Imeida} for a discussion of several minimization algorithms) and in fact is essential if one wishes to discover a minimum partition.
\end{enumerate}
The result of the procedure is a pair of machines,each machine represented as a partitioning of the set of stats into blocks.
\paragraph{}
As mentioned, a procedure exists for finding all possible partitions, corresponding to a two-way decomposition (Lee et al.\cite{Lee}).  So far the best known algorithm for partitioning, which conducts a search of what is called the \textit{partition lattice}, is exponential, in the number of states.  This makes handling large machines infeasible, but we have a remedy to this disagreeable situation which we present later, and for now we ignore this issue in the interest of progress.
\paragraph{}
With a complete list of all possible partitions, it is possible to select from the list a minimal partition.  That is we choose a partition that has the smallest possible number of states.  Based on the theory, this pair of partitions, which we will call $(\pi_I, \pi_D)$, satisfies several properties, other than being minimal.  In fact, these properties are true of any pair of partitions that can be used as a decomposition.
\begin{property}\textit{Sound Decomposition}.
Let $M$ be a machine with no output.  Let $(\pi_I, \pi_D)$ be a pair of partitions of the set $M.V$.  The machine $M$ can be decomposed into two machines, $M_I$ and $M_D$, built from the partitions, $\pi_I$ and $\pi_D$, respectively, if the following hold.
\begin{enumerate}
\item
The two partitions of $M.V$ must be \textit{input-preservative}. We can define input preservation as follows.  Lat $\pi$ be a partition of the set $M.V$.  Let $B \in \pi$ be a block in the partition, and let $u, v \in B$.  Let $(u, i, w_u) \in M.E \wedge (v, i, w_v) \in M.E$.  Then $\exists C \in \pi$ such that $w_u, w_v \in C$.  This property ensures that the partition $\pi$ behaves like $M$, at least to the level of the block.  It is the main driver for the usual FSM minimization algorithm.
\item
The pair of partitions, $(\pi_I, \pi_D)$, are \textit{orthogonal}.  We can define orthogonality in a similar fashion as we do for vectors.  We say that two partitions, $\pi_I$ and $\pi_D$, are orthogonal if and only if $\pi_I \cdot \pi_D = \mathbf{0}$.  This definition uses the dot-product of two partitions, which must also be defined.  Our dot-product is defined as $\pi_I \cdot \pi_D = \bigcup \{B_I \cap B_D\ : B_I \in \pi_I \wedge B_D \in \pi_D\}$.  This property ensures that a block in $\pi_I$, and a block from $\pi_D$ uniquely identify a state in $M.V$.  This, in turn, allows a composite machine to determine, given a sequence of inputs, what state $M$ will be in, and using this information calculate the correct output.
\end{enumerate}
\end{property}
\subsubsection{Constructing Machines from Partitions}
Once we have in our possession the pair of minimal partitions, $(\pi_I,\pi_D)$, of the machine $M$, we need to construct the corresponding finite state machines, $M_I$ and $M_D$.  The easiest of the two machines to construct is the independent machine, $M_I$.
\paragraph{}
In our description of $M_I$, we use the function $e(B)$ to assign a number to each block $B$ in a partition.  That is $e: \pi \rightarrow Z^*$, where $\pi$ is a partition.  We use this numbering of the blocks in a partition to give states numbers that can be used in the independent machine which we are constructing.
\begin{definition}\textit{Independent Machine}. \label{d7} We define the independent machine, for the machine $M$ as\\
\indent
$M_I = (V_I, T_I, M.I, M.I \times V_I, O_I, s_I)$,\\
where $V_I = \{e(B) | B \in \pi_I\}$, $T_I = \{(e(B_1), i, e(B_2)) | \exists (u, i,v) \in M.E \ni u \in B_1 \wedge v \in B_2 \wedge B_1, B_2 \in \pi_I\}$, $O_I = \{(u, i, u) | u \in V_I \wedge i \in M.I\}$, and $s_I = e(B_0)$, with $B_0 \in \pi_I \wedge M.s \in B_0$.
\end{definition}
Summarizing Definition \ref{d7}, we are constructing a machine with transitions that are input preserving, and mimic the original machine $M$ at the block level.  The constructed machine takes the same input as $M$, and outputs the state number of the constructed machine.  This is our typical machine type, already illustrated in Section \ref{s19} for matrix multiplication. 
\paragraph{}
To construct the dependent machine, $M_D$, from the partition $\pi_D$, we use the same scheme we used in Section \ref{s19} for matrix multiplication. We also use the notational aid $\chi : \pi_I \times \pi_D \rightarrow M.V$, a partial function that finds the single common state in two blocks from two orthogonal partitions.  That is, $\chi(B_I, B_D) = v$, where $B_I \in \pi_I \wedge B_D \in \pi_D \wedge v \in B_I \cap B_D$.
\begin{definition}\textit{Dependent Machine}.  We define the dependent machine, for machine $M$ as\\
\indent
$M_D = (V_D, T_D, M.I \times V_I, M.P, O_D, s_D)$,\\
where $V_D = \{e(B) | B \in \pi_D\}$, $T_D = \{(e(B_1), i, e(B_2)) | \exists (u, i,v) \in M.E \ni u \in B_1 \wedge v \in B_2 \wedge B_1, B_2 \in \pi_D\}$, $O_D = \{(u, (i, v), p) | u \in V_D \wedge i \in M.I\wedge v \in V_I \wedge p = \chi(e^{-1}(u), e^{-1}(v)) \}$, and $s_D = e(B_0)$, with $B_0 \in \pi_D \wedge M.s \in B_0$.
\end{definition}
The machine $M_D$ receives the input to $M$, and the current state of $M_I$, and uses these inputs and the function $\chi$ to calculate the current state of $M$.  Knowing these values, it can calculate the output of $M$, and because $M_D$ is input preservative, it can calculate its next state.
\subsubsection{$\mathbf{LPR(\mathit{k})}$: an Extension to the LPR}
The LPR is a fairly simple machine in structure; it is a linear graph.  With function composition, this is not a problem.  The transformation applied to the LPR scrambles it enough to yield a fairly well disguised encrypted machine.  However, when used in machine composition, the simplicity of the machine is a drawback.  To be more direct, if you look for the minimum decomposition of the LPR, you will get the trivial (\textbf{0}, \textbf{1}) pair of partitions.  So, it is imperative that, when doing machine decomposition using the LPR, we extend it into a machine with a more complex topology.
\paragraph{}
Notice that if we extend the LPR carefully, we still have a machine that is derived from the original machine $M$, although with an extra level of indirection.  This is in line with our view of the watermark as an encrypted artifact, as described in Section \ref{p3}.  And, concerning encryption, an advantage of extending the LPR is that this gives us the ability to introduce several keys into the encryption.
\begin{figure} [h]
\centering
\includegraphics[scale=0.25]{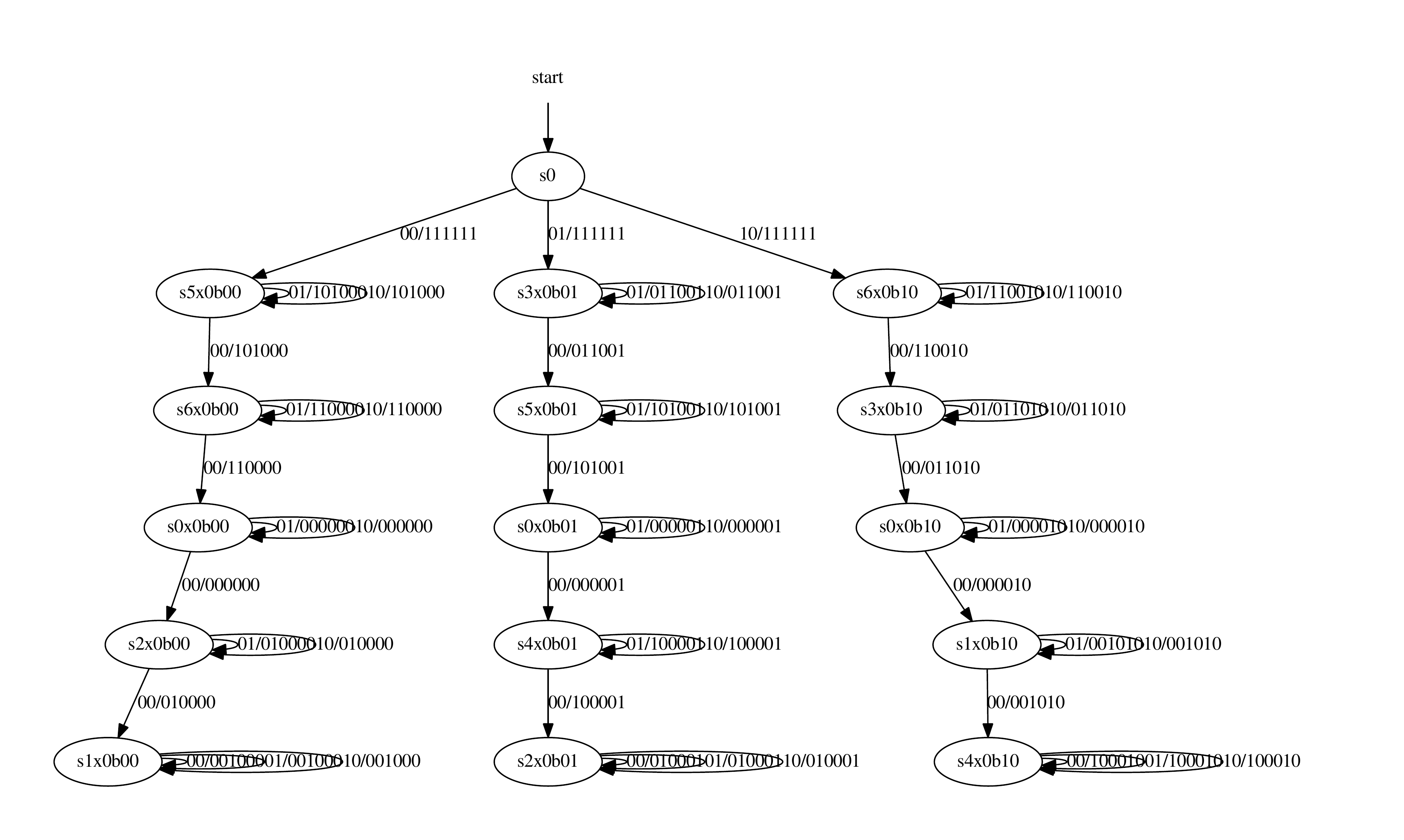}
\caption{Example LPR(3).\label{f14}}
\end{figure}
\paragraph{}
A fairly simple way of extending an LPR is to combine $k$ linear graphs into a single graph called the $LPR(k)$.  For our running example, Figure \ref{f14} shows the $LPR(3)$ graph.  What appears to be happening in this graph is that the LPR has been duplicated three times, and connected to a new start node.  This interpretation is not too far from the truth. However, upon further examination, one realizes that the columns of the $LPR(3)$ are not quite identical, nor is this possible.  Because each state must have a unique state number, although the columns are identical in structure, the states chained together in each column have different state numbers.  To put it another way, each column is not the LPR, but more correctly, a renumbering of the LPR.
\paragraph{}
Obviously, there are many ways to perform a renumbering of the branches of the $LPR(k)$.  The method we use is geared towards simplicity.
\paragraph{}
A renumbering can be thought of as a hash function $h_{r,c} : M.S \rightarrow M.S$.  The values $r$, and $s$ refer to the row and column position of the state in $LPR(k)$.  For instance, the $LPR(3)$ shown in Figure \ref{f14} has a new start state, three columns, and five rows.
\paragraph{}
There are not many requirements on the function $h_{r,c}$.  It is not strictly requisite that $h_{r,c}$ be invertable.  The function used to generate the $LPR(3)$ of Figure \ref{f14} is what we call an 
\textit{add-shift hash function}.  
\begin{definition} \textit{Add-Shift Hash Function}.  Let \\
\indent
$h_{r,c} : M.S \rightarrow M.S$.\\
Further, let $z$ be the size of the binary representation of the states in $M.S$.  That is,\\
\indent
$z = \max\limits_{1 \le j \le \|M.S \|}(\lceil \log(s_j) \rceil : s_j \in M.S)$.\\
Then we define the add-shift hash function $h_{r,c}$ as
\begin{equation}
h_{r,c}(x) = ((x \circlearrowleft c) + r) \mod z,
\end{equation}
where the notation $x \circlearrowleft c$ denotes a binary left rotate of $x$, by $c$ bits.
\end{definition}
That is, in an add-shift hash, the state number is rotated, and an offset is added to it.  Both rotation, and modulus addition are bijections, and so, it turns out, so is our add-shift hash function $h_{r,c}$.  We stress that add-shift hash functions are only one possible choice for renumbering.  Many other possibilities easily come to mind.  In fact, many possibilities that are not quite as regular as add-shift hashing might be superior for encryption purposes.  Since we present the longest path reduction as just a simple example of characteristic machines, the quality of the hash function used in this REDUX is not an issue.
\subsubsection{Final Thoughts on Optimal Machine Decomposition}
\begin{figure} [h]
\centering
\includegraphics[scale=0.25]{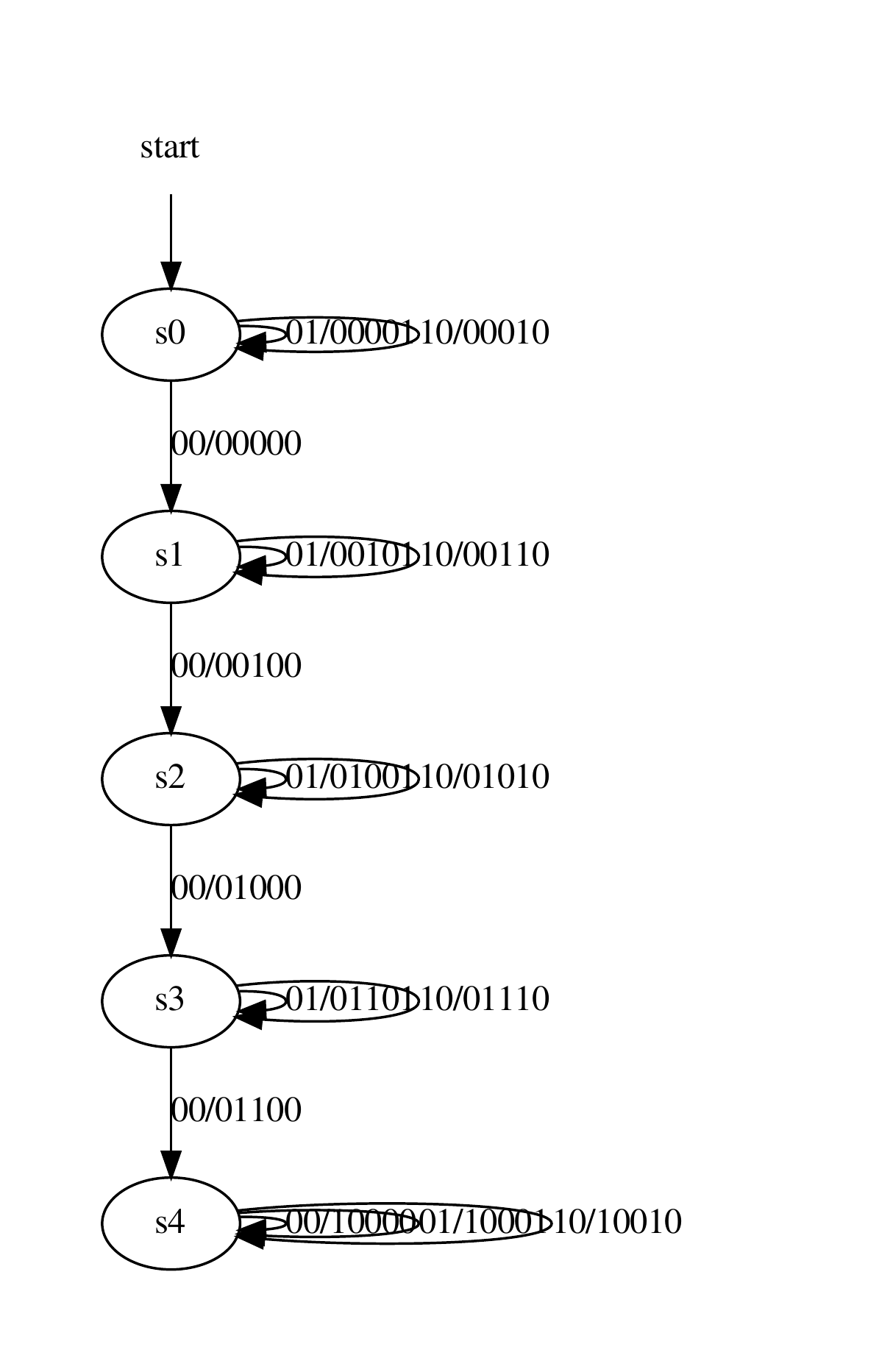}
\caption{Independent machine for the example in Figure \ref{f14}.\label{f33}}
\end{figure}

\begin{figure} [h]
\centering
\includegraphics[scale=0.12]{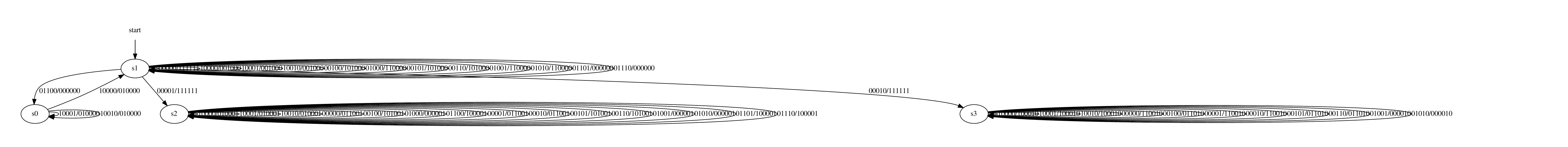}
\caption{Dependnt machine for the example in Figure \ref{f14}.\label{f34}}
\end{figure}

What we have presented so far is a system that produces a decomposition as follows.
\begin{enumerate}
\item
From the machine $M$ calculate a characteristic REDUX.
\item
Enumerate all input-preserving partitions of the REDUX.
\item
Choose a smallest partition, $(\pi_I, \pi_D)$, and build the machines $M_I$, and $M_D$.
\end{enumerate}
\paragraph{}
One problem with doing an optimal decomposition, as mentioned in Section \ref{s22}, is that finding optimal partitions is rather difficult.  A shortcut is to use a partition of a particular, or fixed structure.  Depending on the choice of partition,  this can determine the complexity of the watermark machine, as well as its size.  We examine one such fixed partitioning in Section \ref{s33}.
\paragraph{}
To illustrate our technique, in Figures \ref{f33} and \ref{f34} we have decomposed the $LPR(3)$ of Figure \ref{f14} into an independent machine, and a dependent machine, respectively.  This decomposition was realized by searching the partition lattice of all input-preserving partitions for a decomposition with the smallest total number of states.  Three such partitions were found, and one was presented in this paper.  The partition chosen was \\
\indent
$\pi_I = \{ \{s0, s3x0b01, s6x0b10\}, \{s5x0b00, s5x0b01, s3x0b10\},$\\
\indent
$\;\;   \{s6x0b00, s0x0b01, s0x0b10\}, \{s0x0b00, s4x0b01, s1x0b10\},$\\
\indent
$\;\; \{s1x0b00, s2x0b01, s4x0b10, s2x0b00\} \}$, and\\
\indent
$\pi_D = \{ \{s2x0b00; s0, s5x0b00, s1x0b00, s6x0b00, s0x0b00\},$\\
\indent
$\:\; \{s3x0b01, 5s2x0b01, s5x0b01, s0x0b01, s4x0b01\},$\\
\indent
$\;\; \{s6x0b10, s4x0b10, s3x0b10, s0x0b10, s1x0b10\} \}$.\\
In this decomposition, the partition $\pi_I$ is close to a machine that combines all states in each row of the $LPR(3)$, and $\pi_D$ is very close to a machine that combines all the states in each column of the $LPR(3)$.

\subsection{Approximate Machine Decomposition}\label{s33}
\begin{figure} [h]
\centering
\includegraphics[scale=0.25]{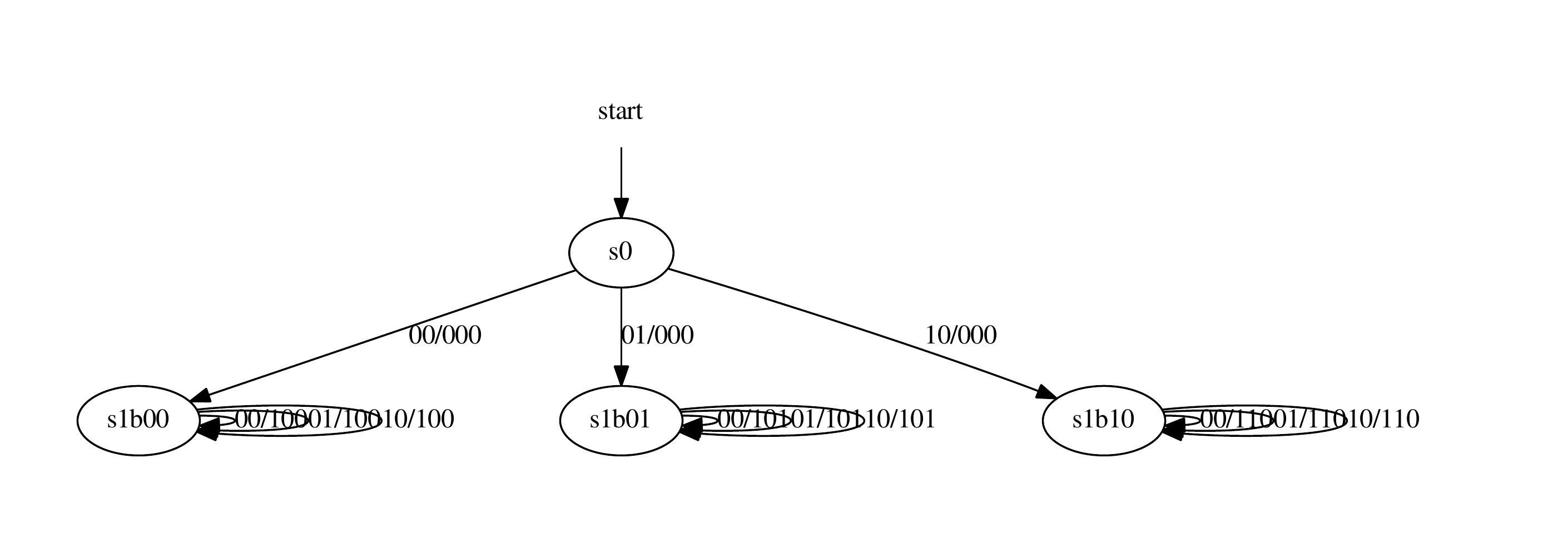}
\caption{Fixed decomposiiton independent machine for  Figure \ref{f14}.\label{f57}}
\end{figure}

\begin{figure} [h]
\centering
\includegraphics[scale=0.25]{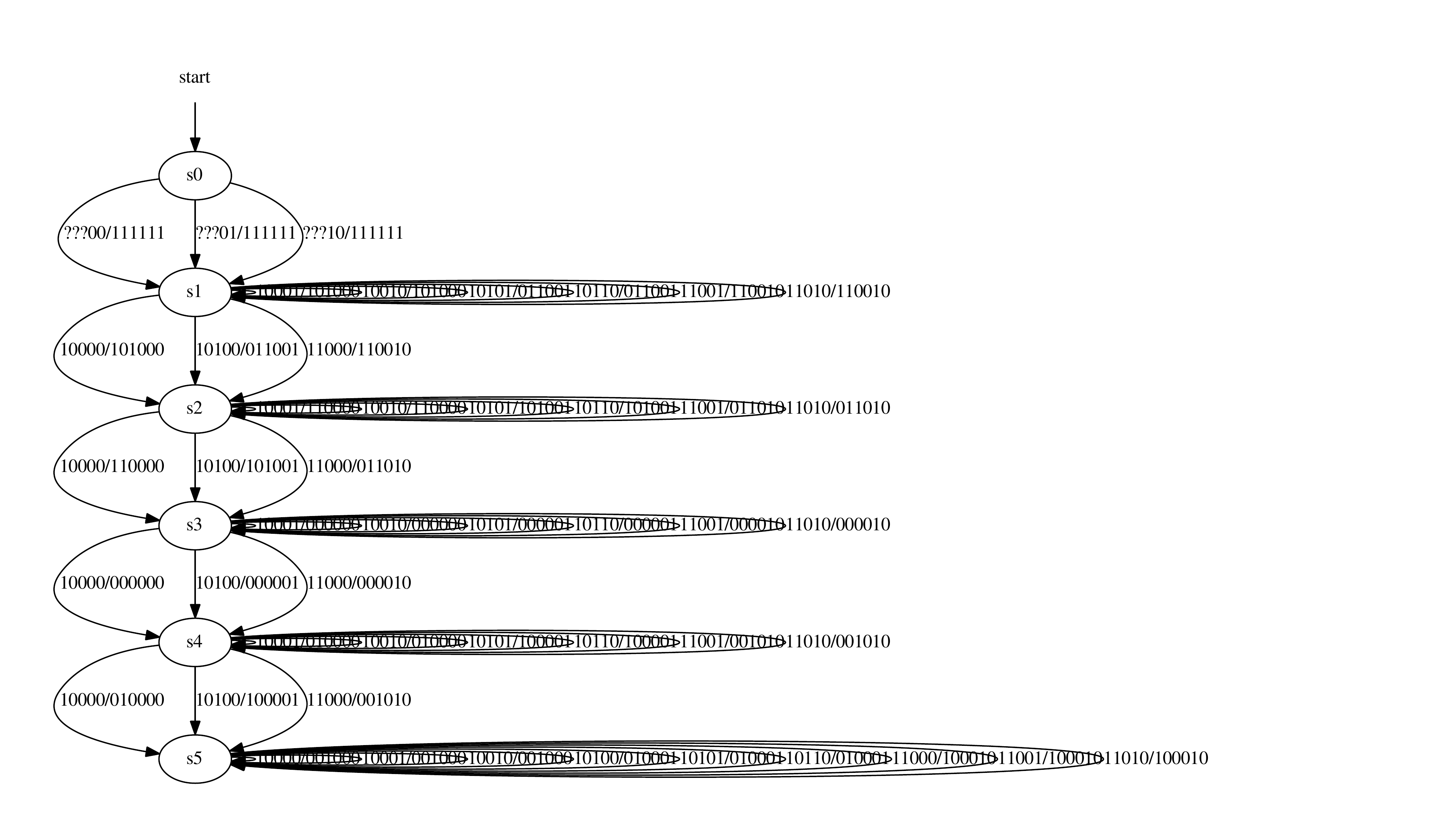}
\caption{Fixed decomposition dependent machine for  Figure \ref{f14}.\label{f58}}
\end{figure}

To avoid a search of the partition lattice, one might be willing to accept a partition that, although not optimal, is very close.  This stratgy is suggested by Lee \& Yannakakis.\cite{Lee}  That is, instead of a search, we would always use a decomposition of a known form, a decomposition that is known to produce machines with near minimal number of states.
\paragraph{}
We often think of a decomposition in terms of spatial coordinate systems.  That is, we can think of the states of a host machine as a collection multi-dimensional points in an Euclidean space, and the decomposition is the process of splinting  the points into several base vectors corresponding to several orthogonal dimensions. Following this line of thought, the $LPR(k)$ has a natural interpretation as a multi-dimensional point.  For example, the states of the $LPR(3)$ can be thought of as uniquely identified by a column number, indicating one of three columns, and a row number, indicating one of five rows.  In general, we can identify the state of an $LPR(k)$ by specifying one of $k$ columns, and one of $\|LPR_M\|$ rows.
\paragraph{}
The previous discussion suggests that we might decompose an $LPR(k)$ into two machines, of a certain form.  The form of these machines is illustrated in Figures \ref{f57} and \ref{f58}, which correspond to the proposed decomposition for the $LPR(3)$ of Figure \ref{f14}.  The independent machine of Figure \ref{f57} branches from a start state to one of three ``column'' states.  The dependent machine of Figure \ref{f58} branches from a start state to a chain of ``row'' states.  The two machines are structured as usual: the independent machine outputs its state number, and the dependent machine outputs the state number of the original $LPR(k)$.
\subsection{Comparison of the Approximate Decomposition to the Optimal Decomposition}
There are obvious advantages of using an approximation over doing a comprehensive search for the optimal solution, in terms of search effort.  However, the question as to the quality of the approximation is of inescapable importance.
\paragraph{}
In our system we have an approximate decomposition of an $LPR(k)$.  We will refer to this decomposition as the \textit{fixed decomposition}.  We are interested in establishing the relationship between the fixed decomposition, and an optimal decomposition.  We do this by first proving a lower bound on the size of an optimal decomposition, and then continue by calculating the size of the fixed decomposition.
\paragraph{}
Our process starts with a machine $M$, and calculates the machine $LPR_M$, and from this calculates $LPR(k)$.  We will assume that $\| LPR(k) \| = n \times k + 1$.  That is the number of states in $LPR(k).V$ is one more than the number of states in a single branch of the $LPR(k)$, $n$, multiplied by the branching factor of the $LPR(k)$, $k$.  We also assume that we decompose the $LPR(k)$ into two machines, $M_I$ and $M_D$, that will be used for the independent machine, and the dependent machine, respectively, in the optimal decomposition.
\begin{lemma}\label{l12}
If $M_I$, and $M_D$ are the decomposition of an $LPR(k)$, then
\begin{equation}\label{e77}
\| M_I \cup M_D \| \ge n + k + 1
\end{equation}
\end{lemma}
\begin{proof}
The proof of this theorem is based, primarily, on the fact that $M_I$, and $M_D$ are orthogonal.  More accurately, 
\begin{equation}
\pi_I \cdot \pi_D = \mathbf{0},
\end{equation}
using the same definition of the dot product of two partitions as Lee et al.\cite{Lee}. If we count only non-empty blocks, then we can say 
\begin{equation}\label{e78}
\|\pi_I \cdot \pi_D \| = \|\mathbf{0}\| = \|LPR(k)\| = n \times k + 1.
\end{equation}
\paragraph{}
Our claim, in Equation \ref{e77}, is that 
\begin{equation}\label{e79}
\|\pi_I\| + \|\pi_D\| \ge n + k + 1.
\end{equation}
Given Equation \ref{e78}, a simple proof by contradiction, based on the size of a cross product of two partitions, can be used to show that Equation \ref{e79} is true, which, in turn, completes the proof of Equation \ref{e77}.
\end{proof}
\paragraph{}
We now have a lower bound on the size of a decomposition.  And, we now look at the size of the fixed decomposition, and compare it with this lower bound.
\begin{lemma}\label{l13}
Let $M_I$, and $M_D$ constitute the fixed decomposition of the machine $LPR(k)$.  The partitions corresponding to the decomposed machines are $(\pi_I, \pi_D)$.  Then
\begin{equation}\label{e81}
\|M_I \cup M_D \| = n + k + 1,
\end{equation}
where $n$ is the branch length of $LPR(k)$, and $k$ is the branch factor of $LPR(k)$.
\end{lemma}
\begin{proof}
The proof is trivial.  Equation \ref{e81} follows from the fact that $LPR(k)$ is a machine with a single start state, and $k$ branches, each branch consisting of $n$ states.
\end{proof}
\paragraph{}
Our two lemmas, Lemma \ref{l12} and Lemma \ref{l13}, can be combined into the following theorem.
\begin{theorem}\label{t56}
If the $LPR(k)$ machine is decomposed into two machines, using the fixed decomposition, then the resulting decomposition is minimal.
\end{theorem}
What Theorem \ref{t56} establishes, while what we began assuming was only an approximation to an optimal decomposition, the fixed decomposition is, in fact, actually an optimal decomposition.  This result, in turn, suggests that if one's goal is simply to use a minimal decomposition, there is no need to perform a search of the partition lattice to find such a decomposition.
\section{Concealing Watermark Outputs}
Our work, so far, describes a distribution package, containing a host circuit, and a watermark circuit.  Each of the two circuits has its own collection of input, and output pins.  What we wish to discuss at this point is the nature of the I/O pins for the watermark circuit.  And, to discuss the nature of these pins, we must consider the types of attacks that the watermark circuit should render infeasable.  We can classify these attacks into two categories, based on the categorization of Abdul-Hamid et al.\cite{Abdelhamid}
\begin{itemize}
\item
\textit{Counterfeit}.  The attacker claims that an artifact belongs to the owner, but it actually belongs to the attacker.
\item
\textit{Appropriation}.  The attacker claims that an artifact belongs to the attacker, but it actually belongs to the owner.
\end{itemize}
There are two other types of claims using the watermark.  We probably would not call these claims attacks, but rather verification.
\begin{itemize}
\item
\textit{Confirmation}. The owner claims the artifact belongs to the owner, and it does.
\item
\textit{Denial}.  The owner claims the artifact does not belong to the owner, and it doesn't.
\end{itemize}
In both scenarios, attack and verification, the watermark circuit should provide strong evidence of the actual status of the artifact.
\paragraph{}
Notice that we are not trying to protect against exact reproduction of a circuit.  This, arguably, would be considered legitimate, and equivalent to taking a photo-copy of a printed document.  It is in the situation where a functional duplicate is being made.  It is assumed that the functional specification is considered the intellectual property of the owner, and is the artifact being protected under patent law.
\subsection{Counterfeit}
In both types of attacks, and both types of verification, a successful defense depends on the ignorance of the attacker, concerning the watermark.
In the case of counterfeiting, the attacker would build a circuit that is functionally equivalent to the host circuit.  The attacker may not include a watermark circuit in the package, if they are not aware that the package contains such a circuit.  This makes it relatively easy for the owner to demonstrate that the counterfeit circuit is not legitimate.
\paragraph{}
If the attacker is aware of the watermark circuit, then they will most likely try to duplicate it.  There might be two approaches to the duplication.  Firstly, the attacker may simply try to discover the behavior of the watermark circuit, and build a circuit that duplicates it.  A second approach is to try to understand, at a high level, the computation that the watermark circuit performs, and build a circuit that does this same computation.  This clearly is the more difficult attack, since the owner is most likely being fairly secretive about the watermark.  In our discussion, we will assume that the owner is unwilling to disclose any information on the watermark circuit, unless required to in litigation.
\paragraph{}
We are left with an attack that involves discovering the behavior of the watermark circuit, in a watermark test.  That is, the attacker seeks to discover, given a sequence of inputs, what are the outputs of watermark circuit.  Notice that in our system, using an $LPR(k)$ as a REDUX, this discovery is complicated, because the attacker must actually discover input-output relationships for several watermark test, as opposed to just one.  This is because the test engineer can arbitrary choose to test any branch of the $LPR(k)$, or for that matter, the may test all branches.
\paragraph{}
To determine the input-output relation of the watermark circuit,  the attacker, first must determine the input and output pins of the package.  This is not to hard, since the data sheet of the device will, no doubt list all pins associated with the host machine, and their functions.  If the attacker is able to determine the input and output pins of the watermark circuit, they can try all possible input sequences, and observe the output.  Notice that the attacker need only try sequences up to size $n$, where $n$ is the branch length of the $LPR(k)$, to explore all possible states of the $LPR(k)$, and determine all transition outputs.
\paragraph{}
To render it difficult to perform this type of trial and error attack, we might try to hide the inputs, and outputs of the watermark circuit.  This has been done by making the outputs side-channel.  Another method is to narrow the bandwidth of the I/O channel for the watermark.   This method transforms the usual parallel communication with the watermark circuit into a serial communication, complicating the task of collecting information from the watermark circuit.  We can collapse the bandwidth of the I/O down to a single pin, giving us a boundary scan test (BST) scenario.  And so, we propose to use a simplified JTAG standard BST to implement the watermark test (see Maunder \& Tulloss\cite{Maunder} for a description of JTAG).
\subsubsection{Incorporating JTAG into Watermarking}\label{s67}
\begin{figure} [h]
\centering
\includegraphics[scale=0.40]{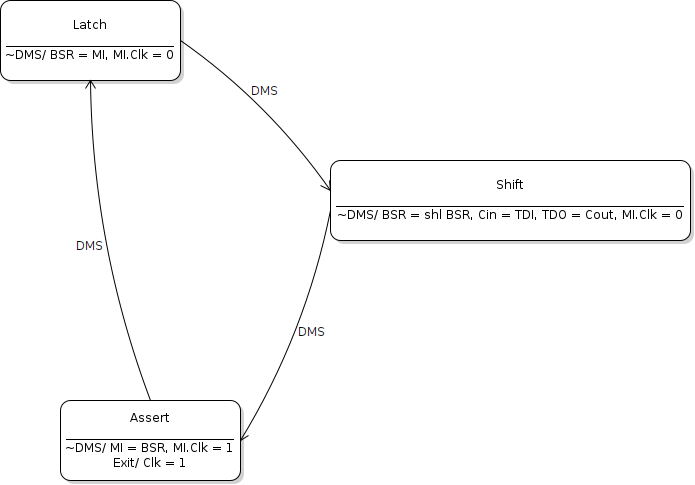}
\caption{TAP machine for simple BST control.\label{f60}}
\end{figure}

In this section we describe how JTAG can be used as an interface to the watermark circuit.  It is not uncommon far a host circuit to be designed for test-ability by including circuitry for BST.  So, it is not unreasonable to assume that this technology is already going to be made accessible in the chip.  The watermark circuit can, therefore, take advantage of the scan chain circuitry, as discussed by Liang et al.\cite{Liang}
\paragraph{}
There are at least two ways to implement a watermark sequence with BST. 
\begin{itemize}
\item
\textit{Combined test and watermark chain}.  The watermark test is incorporated into the existing, standard test scan-chain.  This option is attractive because it makes it appear that there is, in fact, no I/O associated with watermarking.  The attacker must be aware that the BST includes a watermark test, and must discover what part of the BST is the watermark test. 
\item
\textit{Separate test and watermark chains}. The watermark test is on a separate scan chain than the normal circuit test chain.  This option simplifies the job of the test engineer, and allows them to perform a watermark test much more easily.  It is also possible to hide the I/O for the watermark test by using the pins already allocated for the normal scan-chain test.  If done carefully, one extra pin might be all that is needed to switch from host test to watermark test, and vice versa.  
\end{itemize}
In our discussion we will restrict our discussion to separate chains.  This is the option with least design effort. We will describe the pins needed to implement our watermark chain, without attention to whether these pins are used in the normal test chain, or not.  
\paragraph{}
Our BST used for the watermark test is based on JTAG.  The interface consists of four pins.  These four pins control the JTAG test access port (TAP) state machine.
\begin{enumerate}
\item
\textit{Test Clock (TCK)}.  In order to perform a watermark test, this input would be connected to a clock signal, which is used to drive the BST.  This pin would probably be the one extra pin added to the package, and dedicated to the watermark scan.
\item
\textit{Test Mode Selector (TMS)}.  This would be a one bit input that chooses between one of two modes: 0 --- the TAP remains in the same state, and 1 --- the TAP advances to the next state, given the current input value.
\item
\textit{Test Data In (TDI)}.  This is a one bit input that is shifted into the \textit{boundary scan register} (BSR), as input to the TAP.
\item
\textit{Test Data Out (TDO)}.  This is a single bit output that is shifted out of the BSR, as output for the TAP.
\end{enumerate}
\paragraph{}
It is possible to construct a very simple three-state TAP to implement the watermark scan chain.  This machine is shown in Figure \ref{f60}.  In the state \textit{Latch}, the TAP samples the output of the machine $M_I$ and loads it into the BSR, in the state \textit{Shift}, the TAP shifts new input into the BSR, and shifts out new output, and in the state \textit{Assert} the TAP asserts the content of the BSR on to the input lines of $M_I$.  The clock signal to the machine $M_I$ is controlled by the TAP, allowing the TAP to ``freeze'' $M_I$, in order to have time to set up the I/O pins of $M_I$ between trigger edges.
\subsubsection{Permutation of the BST Output}
So far, in Section \ref{s67}, we have described a way of making the managing of the watermark machine quite difficult. In addition to the method described, another technique can be used to render any output of the watermark machine seem random.  This involves permuting the I/O pins of $M_I$, as they are transferred into the BSR.
\paragraph{}
Of course just using a fixed permutation will only transform one watermark sequence into an equally easy to discover different sequence. A second strategy is to vary the permutation used, each time the watermark test is performed.  This can also be done in a fixed pattern, but with similar problems as we have with the fixed permutation.
\paragraph{}
A more resilient approach is rooted in the German Enigma code system (see Hinsley \& Stripp\cite{Hinsley}), where the code varied, depending on a \textit{setting}.  Settings can be thought of as public.  With the enigma machine, knowing a setting was useless, if the attacker was not in possession of the actual machine.
\paragraph{}
In our BST, we are sending the bits of the I/O connections for the $M_I$ to the BSR.  To conceal which bits in the BSR correspond to which I/O bits, we might permute them.  If there are $n$ I/O bits for $M_I$, then there are $n!$ possible permutations. We might choose one of these $n!$ permutations, and consider this our \textit{setting}.
\paragraph{}
To be more precise, let $b \in B^n$ be a bit string, where $B^n = (0 | 1)^n$.  we define a permutation function as follows:
\begin{definition}\textit{Permutation Function}.
We define the function $\rho : B^n \times [1, n!] \rightarrow B^n$ as a permutation function.  The invocation, $d =\rho(b, i)$ performs the $i^{th}$ permutation on the bit string $b$, yielding the bit string $d$.  Conversely, the invocation $b = \rho^{-1}(d, i)$ performs the inverse permutation.
\end{definition}
\paragraph{}
In our scheme, the TAP would have a setting, $i$.  When latching/asserting the BSR from/to the I/O bits for $M_I$, the bits would be permuted using $\rho(b, i)$/$\rho^{-1}(d, i)$, respectively. The interesting part here is that the setting $i$ is not secret, and could be available to the attacker.  What is secret is the function $\rho$, which would be embedded into the TAP design.
\paragraph{}
Returning to our comparison with the enigma machine, the setting for the machine was changed regularly, and the new setting was sent to all users.  We propose that the same could be done with our watermark.  We would propose that each time a watermark test is performed on $M_I$, the setting, $i$, would be changed, randomly.  That is, when a watermark test is initiated, the TAP randomly chooses a setting.  That setting is sent out, through the BSR, unpermuted.  Once all of the bits of the value $i$ have been shifted out of the BSR, then the TAP starts functioning like the TAP described in Section \ref{s67} except that latching and asserting use the functions $\rho$ and $\rho^{-1}$, and the setting $i$, as described in this section.
\paragraph{}
For our proposal to work, the setting must be generated randomly by the TAP.  There ere several choices available for \textit{hardware random number generators} (HRNG).  Most HRNG have a circuit that measures a noise signal, and an ADC to convert the results into a digital form (see Lampert et al.\cite{Lampert} for a discussion).
\paragraph{}
Obviously, incorporating these ``enigma machine'' features into our watermark BST system is going to increase the complexity of the TAP.  The changes, however will make it much harder for an attacker to counterfeit the watermarked circuit.
\subsection{Appropriation}
In an appropriation attack, the attacker claims that the design of the circuit is their design, rather than the design of the actual designer.  To defend against this type of attack, the designer must present design documents.  These design documents can include descriptions of the watermark circuit.  Describing the watermark circuit strengthens the case of the designer.  However, the designer may have to reveal secrets, associated with the watermark in order to make a convincing case. (See Schellekens for a discussion of this issue in Dutch law.\cite{Schellekens}) The strength of such an argument emanates from the fact that the designer can demonstrate that the watermark circuit is not just arbitrary.  This is information that a the designer knows, and the attacker most likely does not.
\paragraph{}
The above discussion, brings to light a problem with watermarking.  To use a watermark,  you have to reveal it.  Once you have revealed the existence and structure of a watermark.  It is rendered less effective, and in future artifacts the watermark may have to be altered.
\section{Security Analysis}\label{sSum}
In this article we have presented a system for watermarking a circuit,packaged as an IC.  To be more precise, we have presented a range of watermarking measures to use in circuitry: form very simple watermarking, but easily attacked, to a complex procedure, that is more difficult to attack.  
\paragraph{}
A salient feature of our watermarking is that we do not use side-channel techniques.  The watermark circuit, although not overt, is visible to an attacker.  The resilience of the watermark is then dependent solely on how easy it is for the attacker to discover secrets buried in the watermark circuit, and make sense of that which is visible.
\begin{figure} [h]
\centering
\includegraphics[scale=0.50]{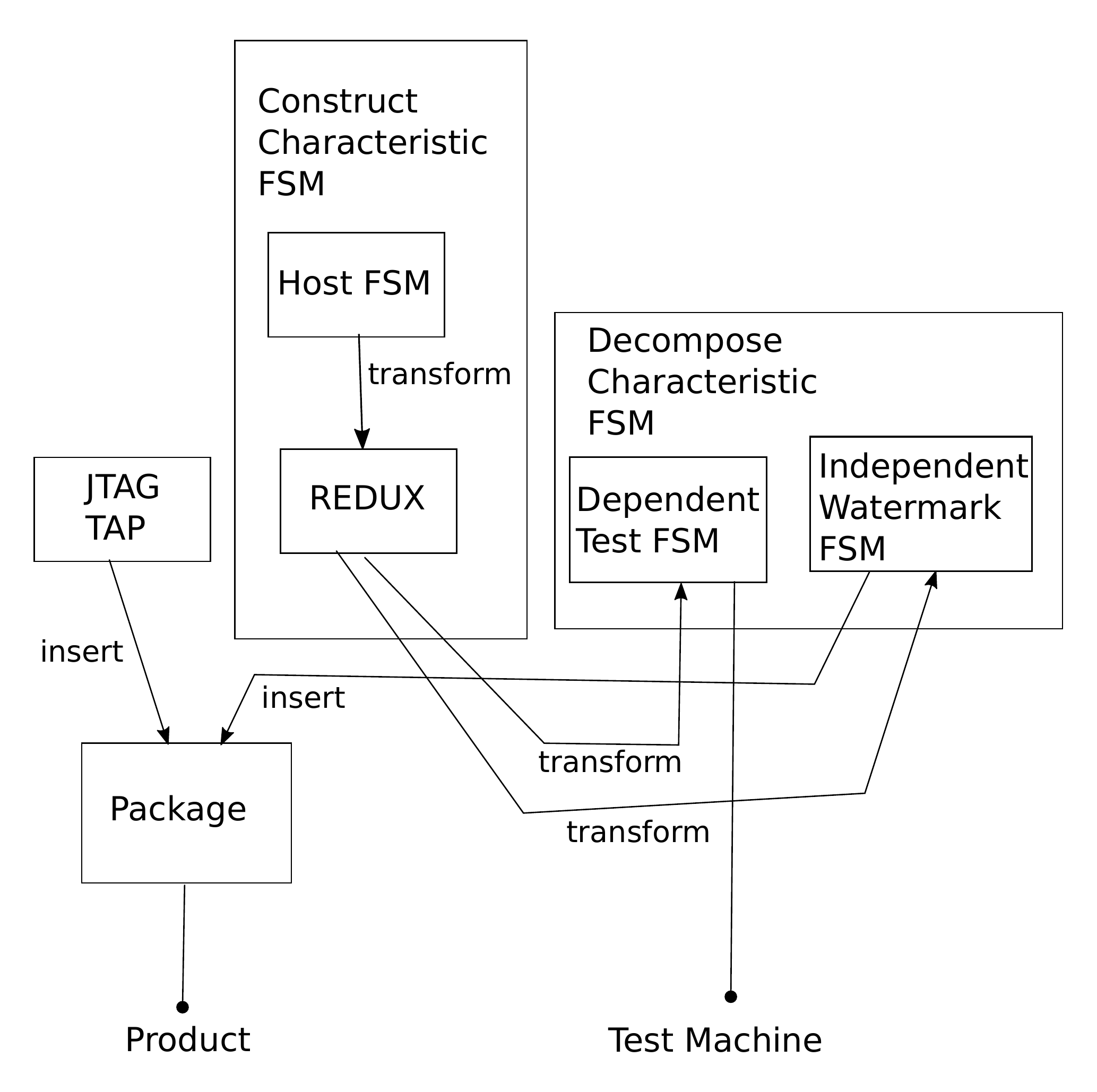}
\caption{Components of the watermarking process.\label{fDC}}
\end{figure}
\subsection{Components}
The main components to our watermark are the following, which are illustrated in Figure \ref{fDC}.
\begin{enumerate}
\item
\textit{A Characteristic Machine}.  From the original machine being watermarked, we derive a characteristic machine, a REDUX, that is dependent on the structure of the host machine.  The nature of the dependence is a secret.  In our work we suggest using the longest path reduction, LPR, of the host machine as the characteristic machine, but this choice is not requisite.
\item
\textit{Decomposition of the Characteristic Machine}.  The characteristic machine is decomposed using a cascade decomposition into a dependent, and independent machine.  The independent machine is used as the watermark circuit.  This decomposition tends to hide the nature of the characteristic machine, protecting this secret.
\item
\textit{Introduction of Serial Communication}. The natural parallel communication with the watermark FSM is replaced by a serial communication system.  A serial protocol is introduced.  This protocol is a secret.  In fact it is suggested that the protocol have elements of randomness, and non-repeatability in it, making the protocol much harder to analyze. The serialization of the watermark machine communication renders the watermark machine much more resilient to duplication and analysis.
\end{enumerate}
The three components to our watermarking, protect the nature of the characteristic machine, as well as protecting the nature of the independent machine in the decomposition.
\subsection{Efficacy}
We now consider the effectiveness of our watermark process. In our discussion, we are concerned with the ease with which an attacker can determine the nature of the watermark machine.  We start by assuming that the attacker has no prior knowledge of element of the watermarking procedure. Therefore, the full process is secret, and the attacker must deduce the structure of the watermark machine from the observable interaction with the input and output ports of an IC. Later we relax our assumption of full ignorance, and consider cases involving more informed attacks.
\paragraph{}
We will further assume that all levels of Figure \ref{fDC} have been implemented: the characteristic machine is the $LPR(k)$, the fixed decomposition has been used to build a minimal dependent, and independent machines, a scan-chain has been introduced for watermark testing, and an HRNG has been built to permute the scan-chain output.
\subsubsection{Cracking the JTAG protocol}\label{sjtag}
With the assumptions we have made, the attacker is faced with a single bit output, TDO, and a single bit input, TDI, in addition to the I/O ports of the host machine.  We will assume that the attacker is aware that a JTAG type protocol has been implemented, for BST, and that there is a watermark test incorporated in the JTAG testing system. 
\paragraph{}
To determine the correct function of the watermark machine, the attacker must be able to determine all sequences of inputs that will cause a correct test, and all correct output sequences, resulting from the correct tests.  To perform this task, the attacker must first decipher the format of the JTAG protocol used.
\paragraph{} 
To crack the operation of the JTAG protocol, the attacker must essentially determine the behavior of an FSM that is a combination of the TAP, and the independent machine.  For this machine, the attacker must discover the full watermark sequence, without knowing anything about the underlying FSM. This is a difficult problem.  It involves, firstly, determining when a watermark sequence ends.  That is, the attacker must determine the length of the watermark sequence.
\paragraph{}
If the designer is watermarking the host machine using the fixed decomposition, the independent machine is a multi-branch machine.  As such, the machine produces a collection of outputs, which depends on the number of branches in $LPR(k)$  So, one property of the dependent machine that must be discovered is how many branches, $k$, are in $LPR(k)$ 
\paragraph{}
We can rephrase this problem.  Determining the size of the collection of outputs in the dependent machine, is equivalent to determining how many columns are in $LPR(k)$, by simply manipulating inputs, and observing the output of the independent machine with a TAP.  The difficulty of the general form of this problem is  established in the following theorem.
\begin{theorem}\label{TP}
Let \emph{M} be an FSM, with a single output bit, and a single input bit.  Further assume that the output of the machine is available, and the input of the machine can be manipulated, but no information concerning the structure of the machine is available.  Then, the problem of determining the number of outputs produced by the REDUX is undecidable.
\end{theorem}
\begin{proof}
It is not to difficult to prove that the number of outputs produced by any machine $M$ cannot be determined algorithmically, by observing its response to input.  The proof is by contradiction.  It hinges on the fact that if you stop analysis of the FSM $M$ after a finite time, after seeing $j$ outputs, and declare the number of outputs as $j$, then it is possible to construct a machine $M'$ that produces the same observed behavior as $M$, but has more than $j$ outputs.  Therefor, it is impossible to determine the number of outputs with certainty.
\end{proof}
\paragraph{}
The above theorem establishes that, even if the attacker knows that the watermark machine is multi-branch linear, they will be unable, with complete certainty, to establish its size.  This theorem, however, does not preclude a heuristic approximation.
\paragraph{}
In the attacker's analysis, the fact that the machine $M$ is using a BST interface is not intentionally hidden,  However, the BST protocol hides the number of I/O pins of the independent machine, which is an important secret.
\subsubsection{Cracking an Unprotected Independent Machine}\label{sim}
As mentioned in Section \ref{sSum}, a range of protection can be placed on the watermark circuitry.  In this section we examine the security of the independent machine of the fixed decomposition, without being hidden with a scan chain interface.  In this case, the circuit package presents several parallel output pins, and several parallel input pins.
\paragraph{}
Recall that the input to the independent machine is the same input received by the REDUX.  As we are assuming that the attacker has access to the parallel output, and input of the independent machine, then the attacker knows the sizes of the input and output, in terms of number of bits.  We will assume, however, that the attacker knows nothing about the structure of the independent FSM.
\paragraph{}
The sizes of the input, and output of the independent FSM are useful information for the attacker.  Initially, it looks like this information gives the attacker almost complete knowledge of the size of the independent machine.  Knowing the size of the input certainly is sufficient to deduce the number of possible inputs.  Knowing the size of the output would appear to inform the attacker as to the number of states in the independent machine.  But, this is not the case.  Knowing the size of the output allows the attacker to determine the number of possible outputs, but this is different from the number of states.  The number of states in the independent machine still remains hidden, leading to the following theorem.
\begin{theorem}\label{tup}
Let \emph{M} be an FSM, with \emph{m} input bits, and \emph{n} output bits.  Assume that the output of the \emph{M} is available, and that the input bits can be manipulated, but no information is available concerning the structure of the machine.  Then determining the number of states and transitions composing \emph{M} is an undecidable problem.
\end{theorem}
\begin{proof}
The proof of the theorem follows from our above observation that nothing is know about the number of states, nor can this information be deduced.  Then, a similar proof sketched out for Theorem \ref{TP}, can be applied to complete the proof of Theorem \ref{tup}.
\end{proof}
\paragraph{}
Theorem \ref{tup} establishes that the problem of cracking the independent machine is difficult.  It is closely related to Theorem \ref{TP}.  That is, the two theorems establish that it is difficult to deduce the internal structure on an FSM just by observing its input and output.
\subsubsection{A more Informed Attack}\label{sIA}
In Sections \ref{sjtag} and \ref{sim} we see that, if nothing is known about the internal structure of the independent machine, that structure is difficult to discover with certainty.  It was pointed out that a heuristic approximation to the independent machine is possible, but then the attacker is never quite sure that the approximation will be able to pass as the actual watermark machine.
\paragraph{}
It is also quite possible that the attacker does, in fact, know something about the structure of the independent machine.  In this section we discuss how this might affect the efficacy of the attack.
\paragraph{}
For the purposes of this discussion, we will assume that the attacker is aware that the watermark machine is an $LPR(k)$. Further the attacker knows that the $LPR(k)$ has been decomposed using the fixed decomposition.  In that case, the attacker is aware that the independent machine is a machine similar to that in Figure \ref{f57}, with multiple branches from a start state.  As in Section \ref{sim}, we also assume that the attacker has access to the input and output of the independent machine.  
\paragraph{}
The amount of information in possession of the attacker is significant, with the given assumptions.  Actually, the attacker knows enough information to take a normally undecidable problem, and render it decidable.
\paragraph{}
Perhaps, initially, it looks as if the only secret denied to the attacker is the branching factor of the $LPR(k)$, $k$.  But after a little thought one realizes that the attacker knows a bound to this value also.  This realization leads to the following theorem.
\begin{theorem}\label{ttt}
Let \emph{M} be a fixed decomposition of the \emph{LPR(k)} of a host machine.  Assume that the output of \emph{M} is available, and the input is likewise available.  Further assume that the fact that \emph{M} is a fixed decomposition of an \emph{LPR(k)} is not secret.  Then, a machine \emph{M'} that is equivalent to \emph{M} can be constructed algorithmically in time which is \emph{O}$(2^\chi)$, where $\chi$ is the size an input in bits.
\end{theorem}
\begin{proof}
The proof is constructive.  We sketch the process of constructing the machine $M'$.  To construct $M'$ we trace the operation of the machine $M$, starting from the start state.  In the start state, we try all $2^\chi$ possible input values.  We follow each of these input trials by trying any arbitrary input value and observing the output.  This output tells us what branch state we have arrived in, and in this way we discover all $k$ legitimate branches of the machine $M$, and all states of the machine $M$, where $k \le 2^\chi$.  The time complexity of $O(2^\chi)$ follows from the number of input value that we tried.
\end{proof}
\paragraph{}
Theorem \ref{ttt} demonstrates a not too surprising result: the more information we know about the watermarking process, the easier it is to analyze the independent machine.  This is one reason to use an optimum decomposition rather than a fixed decomposition.  With an optimum decomposition, the structure of the the independent machine can be altered, and varied, to a certain degree,making the machine less susceptible to an informed attack.
\subsection{Constructing the Watermark Machine from the Independent Machine}
In this section, we assume that an attacker has been able to successfully construct an independent machine, in an appropriation attack.  As a final defense, the licenser might ask the attacker to show why the independent machine was used as a watermark, In this case, the strongest response by the attacker would be to produce the full REDUX from the independent machine, and explain how the REDUX was derived from the host machine.  This response requires that the attacker be able to perform two tasks:
\begin{enumerate}
\item \label{B1}
From independent machine, construct the full REDUX, and
\item \label{B2}
Show how the $LPR(k)$ corresponds to the host machine.
\end{enumerate}
\subsubsection{Reconstructing the REDUX}
To illustrate the complexity of the problem, we assume that the REDUX is an $LPR(k)$.  As in Section \ref{sIA}, Bullet \ref{B1}, we will assume that, since the attacker was able to construct a reasonable facsimile of the independent machine, the attack knows what the independent machine looks like, as well as knowing that the watermark machine is the $LPR(k)$.  The attacker, however, although aware of the function of the host  machine, does not know the structure of the host machine, upon which the $LPR(k)$ is based.
\paragraph{}
More specifically, the attacker knows the number of columns, $k$, in the $LPR(k)$.  There are several things that the attacker, however, is not aware of.  One secret the attacker does not possess is the knowledge of how many rows, $m$, are in the $LPR(k)$.  
\paragraph{}
The attacker might approach the problem of finding the number of rows in the dependent machine by guessing, but they have no method to gauge whether any of the guesses they make actually duplicate the output of the dependent machine.  In other words, the attacker is looking for a goal, while knowing nothing about the goal machine, other than it is linear.  
\paragraph{}
The problem here is that there is an unbounded number of dependent machines that can be used with the independent machine.  Each one implements one of an unbounded number of composed machines, one of which is the $LPR(k)$.  And, none of them are disallowed by the given dependent machine.  This leads to the following theorem.
\begin{theorem}\label{tun}
Given complete knowledge of an independent machine only, the dependent machine, as well as the composed machine, are undefined.
\end{theorem}
Theorem \ref{tun} establishes that given knowledge of the independent machine, the attacker cannot even come close to guessing the structure of the dependent machine.  We have shown this when the attacker knows the REDUX is an $LPR(k)$, and of course, it is true when the attacker is faced with the harder problem in which no information about the REDUX is available.
\subsubsection{Partial Mapping of an $LPR(k)$ to a Host Machine}
As a final consideration,  we look at showing that an $LPR(k)$ corresponds to a particular host machine, as mentioned in Section \ref{sIA}, Bullet \ref{B2}.  In this section we assume that the attacker has, somehow, been able to determine the structure of the $LPR(k)$.  That is, the attacker knows the number of rows, $m$, and the number of columns, $k$, of the $LPR(k)$.  Then, the question we ask is how hard is it for the attacker to determine the output of each of the transitions in the $LPR(k)$?
\paragraph{}
With no information on the structure of the host FSM, the attacker will have difficulty determining the output of the dependent machine.
\begin{theorem}
Assume an attacker has full knowledge of the dependent machine, knows the number of rows, and columns in the $LPR(k)$ watermark machine, but knows nothing about the number of states and the transitions of the host machine.  Then, the output of the $LPR(k)$ is uncomputable.
\end{theorem}
\begin{proof}
The proof of this theorem is similar to that of Theorem \ref{tun}.  Without knowledge of the state numbers of the host FSM, and the connection graph, there are an unbounded number of machines that could have an LPR of the same length as the number of rows in $LPR(k)$, each generating a different sequence of state numbers.
\end{proof}
\subsection{Summary}
What we have seen in this section is that our system of watermarking is hard to crack, if certain information can be kept secret.  The situation can be summarized as follows.
\begin{itemize}
\item
if a watermark is protected by a scan chain one may be able to attack it heuristically, but an algorithmic solution is impossible.
\item
If a watermark is unprotected by a scan-chain, but the type of decomposition is unknown, again it is impossible to crack it, with certainty.
\item
If the structure of the independent machine is somehow discovered, there is still no way of determining the structure of the full watermark machine.
\end{itemize}
\paragraph{}
Given the above bullets, we see that the scan-chain protection is very useful, and that the partition used in the decomposition must, at all costs be kept secret.  This precludes always using a single type of fixed decomposition. 
\paragraph{}
We use the LPR as the characteristic machine for the host machine, and then use the $LPR(k)$ as the watermark machine.  Our findings do not necessarily preclude using the $LPR(k)$ always, but this would be risky.  If this fact is discovered by an attacker, it will help them glean information about the structure of the partitions used for decomposition.
\section{Conclusion}
We have compared the protection of a watermark circuit to the encryption of an artifact.  In our scheme circuits are ``encrypted'' using a symmetric key.  Decryption machines can then be constructed, either in hardware or software, to decrypt the output of an encrypted machine, resulting in a full watermark machine, which makes sense of the output for a test engineer.
\paragraph{}
The paper presents three levels of encryption for the watermark machine.  These levels are implemented as three different  techniques.  
\begin{enumerate}
\item
\textit{Matrix Multiplication}.  A technique is presented based on matrix multiplication.  This technique is considered to be the least effective encryption method, but is useful in low risk situations.
\item
\textit{Fixed Decomposition}.  This technique is based on a decomposition with a standard structure.  The decomposition is more complex than the matrix multiplication technique, and so is considered more effective in hiding the nature of the watermark machine.
\item
\textit{Minimum Decomposition}.  In this technique, a minimum decomposition of the watermark machine is used.  Because there are most often several minimum decompositions, this technique gives the circuit owner the flexibility of choosing among the candidates, and potentially introduces a secret, in terms of the structure of the decomposed machine.  Although it is shown that the fixed decomposition technique produces a minimal decomposition,  the minimum decomposition method is more general, and introduces an additional element of secrecy, and so is considered more effective at protecting the watermark machine.
\end{enumerate}
\paragraph{}
Our work has been implemented in simulation, as well as in FPGA.  This includes the TAP for the JTAG feature.  The random assert/latch permutation, using the HRNG has not been implemented, and although the design is complete we are currently working on the implementation.
\paragraph{}
Our system minimizes the circuitry added as the watermark.  It is necessary to add circuitry in order to add any watermark,  but, as in our system, the design goal should be to add enough circuitry to make the watermark effective, but add little enough so as not to significantly effect the performance of the host circuit.  We believe we have achieved this goal, to a great extent.

\section* {Authors}
\begin{description}
	\item{\textbf{James Gil de Lamadrid}}
	is a professor of computer science at Bowie State University.  He has a PhD. from the University of Minnesota, and specializes in programming languages and computer organization.
	\item{\textbf{Seonho Choi}} is a professor of computer science at Bowie State University.  He has a PhD. from the University of Maryland, and specializes in Computer Networks.
	
\end{description}

\end{document}